\documentclass[twoside, twocolumn]{article}     

\usepackage[american]{babel} 
\usepackage[T1]{fontenc}
\usepackage{times}
\usepackage[pdftex]{graphics}
\usepackage{amsmath}    
\usepackage{amssymb}

\graphicspath{{figures/}}

\usepackage{natbib} 
\bibpunct[, ]{(}{)}{;}{a}{}{,}

\usepackage[preprint]{amsstyle}
\newcommand  {\about} {\mathop{\sim}\!}
\newcommand  {\divh}  {\vec{\nabla}_h \cdot}
\newcommand  {\divy}  {\mathop{\mathrm{div}}}
\newcommand  {\curl}  {\vec{\nabla}_h \times}
\renewcommand{\div}   {\vec{\nabla} \cdot}
\newcommand  {\gradh} {\vec{\nabla}_h}
\newcommand  {\grad}  {\vec{\nabla}}
\renewcommand{\<}     {\langle}
\renewcommand{\>}     {\rangle}

\reference{\textit{Journal of the Atmospheric Sciences} (in press)}
\date{Submitted 28 March 2008; in final form 24 September 2008}

\begin{document}

\title{Formation of Jets and Equatorial Superrotation on Jupiter}

\author{Tapio Schneider%
  \thanks{\emph{Corresponding author address:} Tapio Schneider,
    California Institute of Technology, 1200 E.\ California Blvd.,
    Pasadena, CA 91125-2300.  E-mail: tapio@caltech.edu}~
  and Junjun Liu}%
\address{California Institute of Technology, Pasadena, California}

\AbstractText{%
  The zonal flow in Jupiter's upper troposphere is organized into
  alternating retrograde and prograde jets, with a prograde
  (superrotating) jet at the equator. Existing models posit as the
  driver of the flow either differential radiative heating of the
  atmosphere or intrinsic heat fluxes emanating from the deep
  interior; however, they do not reproduce all large-scale features of
  Jupiter's jets and thermal structure. Here it is shown that the
  difficulties in accounting for Jupiter's jets and thermal structure
  resolve if the effects of differential radiative heating and
  intrinsic heat fluxes are considered together, and if
  upper-tropospheric dynamics are linked to a magnetohydrodynamic
  (MHD) drag that acts deep in the atmosphere and affects the zonal
  flow away from but not near the equator. Baroclinic eddies generated
  by differential radiative heating can account for the off-equatorial
  jets; meridionally propagating equatorial Rossby waves generated by
  intrinsic convective heat fluxes can account for the equatorial
  superrotation. The zonal flow extends deeply into the atmosphere,
  with its speed changing with depth, away from the equator up to
  depths at which the MHD drag acts. The theory is supported by
  simulations with an energetically consistent general circulation
  model of Jupiter's outer atmosphere. A simulation that incorporates
  differential radiative heating and intrinsic heat fluxes reproduces
  Jupiter's observed jets and thermal structure and makes testable
  predictions about as-yet unobserved aspects thereof. A control
  simulation that incorporates only differential radiative heating but
  not intrinsic heat fluxes produces off-equatorial jets but no
  equatorial superrotation; another control simulation that
  incorporates only intrinsic heat fluxes but not differential
  radiative heating produces equatorial superrotation but no
  off-equatorial jets. The proposed mechanisms for the formation of
  jets and equatorial superrotation likely act in the atmospheres of
  all giant planets.}

\maketitle

\section{Introduction}

\begin{figure}[!htb]
  \centering\includegraphics{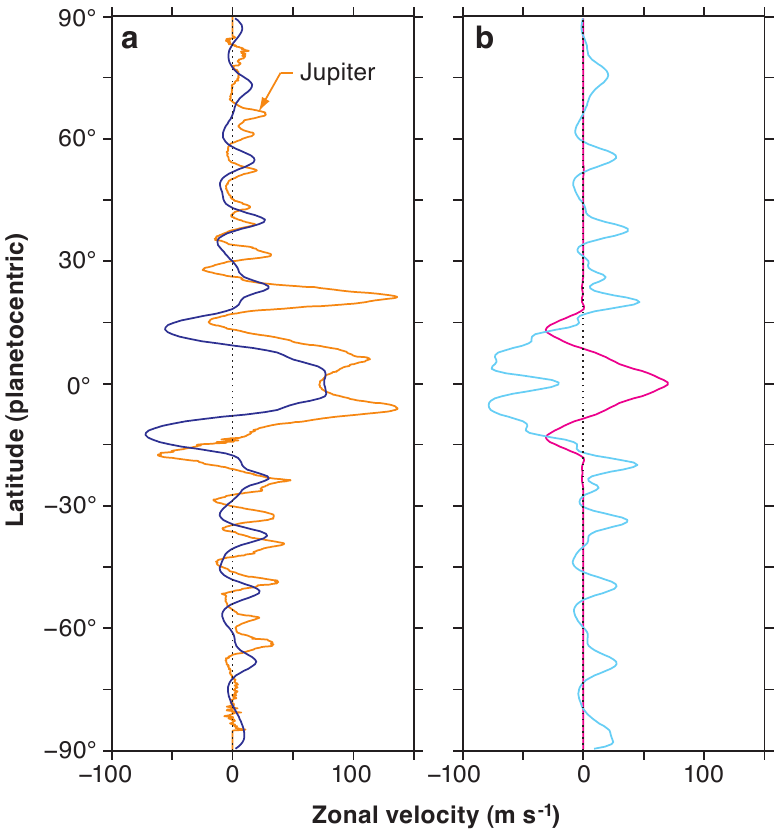}
  \caption{Zonal flow in Jupiter's upper troposphere and in
    simulations. (a) Zonal velocity on Jupiter, inferred by tracking
    cloud features from the Cassini spacecraft \citep{Porco03}
    (orange) and in Jupiter simulation at 0.65~bar (blue). (b) Zonal
    velocity at 0.65~bar in control simulations: with intrinsic heat
    fluxes but with uniform insolation at the top of the atmosphere
    (magenta), and with differential insolation but without intrinsic
    heat fluxes (light blue).  Zonal velocities in simulations are
    zonal and temporal means in statistically steady states (over 1500
    days for Jupiter simulation and over 900 days for control
    simulations); differences between the (statistically identical)
    hemispheres here and in subsequent figures are indicative of
    sampling variability. For Jupiter, latitude here and throughout
    this paper is planetocentric; the simulated planets are spherical,
    so planetocentric and planetographic latitudes are identical.}
  \label{f:winds}
\end{figure}

The zonal flow in Jupiter's upper troposphere has been inferred by
tracking cloud features, which move with the horizontal flow in the
layer between about 0.5 and 1~bar atmospheric pressure
\citep{Ingersoll04,West04,Vasavada05}. In this layer, the zonal flow
is organized into a strong prograde (superrotating) equatorial jet and
an alternating sequence of retrograde and prograde off-equatorial jets
(Fig.~\ref{f:winds}a). This flow pattern has been stable at least
between the observations by the Voyager and Cassini spacecrafts in
1979 and 2000, with some variations in jet speeds, for example, a
slowing of the prograde jet at $21^\circ$N planetocentric latitude by
$\about 40\,\mathrm{m\,s^{-1}}$ \citep{Porco03,Ingersoll04}. The zonal
flow in layers above the clouds has been inferred from the thermal
structure of the atmosphere, using the thermal wind relation between
meridional temperature gradients and vertical shears of the zonal
flow. Meridional temperature gradients and thus vertical shears
between 0.1 and 0.5~bar are generally small: meridional temperature
contrasts along isobars do not exceed $\about 10\,\mathrm{K}$
\citep{Conrath98,Simon-Miller06}. The thermal stratification in the
same layer is statically stable \citep{Simon-Miller06,Read06}.  About
the zonal flow in lower layers, it is only known that at one site at
$6.4^\circ$N planetocentric latitude, where the Galileo probe
descended into Jupiter's atmosphere, it is prograde and increases with
depth from $\about 90\,\mathrm{m\,s^{-1}}$ at 0.7~bar to $\about
170\,\mathrm{m\,s^{-1}}$ at 4~bar; beneath, it is relatively constant
up to at least $\about 20$~bar \citep{Atkinson98}. At the same site,
the thermal stratification is statically stable but approaches
neutrality with increasing depth between 0.5 and 1.7~bar; beneath, it
is statically nearly neutral or neutral up to at least $\about 20$~bar
\citep{Magalhaes02}. These are the large-scale features (if the
Galileo probe data are representative of large scales) that a minimal
model of Jupiter's general circulation should be able to reproduce.

There are two plausible energy sources for Jupiter's general
circulation. First, $\about 8\,\mathrm{W\, m^{-2}}$ of solar radiation
are absorbed in Jupiter's atmosphere \citep{Hanel81}, where the
time-mean insolation at the top of the atmosphere, given Jupiter's
small obliquity of $3^\circ$, varies approximately with the cosine of
latitude. Second, $\about 6\,\mathrm{W\, m^{-2}}$ of intrinsic heat
fluxes emanate from Jupiter's deep interior
\citep{Ingersoll04,Guillot04,Guillot05}; observations of convective
storms \citep{Gierasch00,Porco03,Sanchez-Lavega08} and the neutral or
nearly neutral thermal stratification along much of the Galileo probe
descent path show that the intrinsic heat fluxes are at least
partially convective. Existing models of Jupiter's zonal flow posit as
the driver of the general circulation either differential radiative
heating of the atmosphere or intrinsic convective heat fluxes
\citep[e.g.,][]{Busse76,Busse94,Williams79,Ingersoll04,Vasavada05};
however, they do not reproduce all large-scale features of the jets
and thermal structure. For example, in parameter regimes relevant for
Jupiter, they generally do not produce equatorial superrotation,
unless artifices are employed such as imposing an additional heat
source near the equator \citep{Williams03a} or assuming excessively
viscous flow and permitting intrinsic heat fluxes several orders of
magnitude stronger than Jupiter's \citep{Heimpel05,Heimpel07}.

Here we show that the difficulties in accounting for Jupiter's jets
and thermal structure resolve if the effects of differential radiative
heating and intrinsic heat fluxes are considered together, and if
upper-tropospheric dynamics are linked to drag that acts deep in the
atmosphere and affects the zonal flow away from but not near the
equator. The key is to distinguish the different ways in which eddies
can be generated near the equator and away from it, and to consider
their role and the role of drag at depth in the balance of angular
momentum (the angular momentum component about the planet's spin
axis).

First we describe how eddies can be generated near the equator and
away from it, how convectively generated equatorial waves can lead to
equatorial superrotation, and how the angular momentum balance of the
upper troposphere is linked to drag and constrains the flow at depth
(sections~\ref{s:am_fluxes}--\ref{s:deep_flow}). Then we use
simulations with a three-dimensional general circulation model (GCM)
of a thin shell in Jupiter's outer atmosphere, with an idealized
representation of effects of drag deep in the atmosphere, to
demonstrate that the mechanisms proposed can account for Jupiter's
observed jets and thermal structure
(sections~\ref{s:gcm}--\ref{s:controls}).

\section{Eddy generation and angular momentum
  fluxes}\label{s:am_fluxes}

Tracking of cloud features shows that eddies in Jupiter's upper
troposphere transport angular momentum meridionally from retrograde
into prograde jets; the eddy angular momentum fluxes extend at least
over the layer between about 0.5 and 1~bar, and imply there a mean
conversion rate from eddy to mean flow kinetic energy per unit mass of
order $10^{-5}$ to $10^{-4}\,\mathrm{W\,kg^{-1}}$
\citep{Ingersoll81,Salyk06}. If the eddy angular momentum fluxes
extended unabatedly over a layer more than tens of bars thick, the
total conversion rate from eddy to mean flow kinetic energy would
exceed the rate at which the atmosphere takes up energy from solar
radiation and intrinsic heat fluxes combined. For example, if the eddy
angular momentum fluxes extended over a layer of 2.5~bar thickness,
and if vertical variations of the zonal flow over this layer are
negligible, the total conversion rate would already be of order
$1\,\mathrm{W\,m^{-2}}$---that is, it would amount to $\about 5\%$ of
the energy uptake by the atmosphere \citep{Salyk06}. But probably only
a small fraction of the atmosphere's energy uptake is available to
generate eddy kinetic energy. This means that columnar-flow models of
Jupiter's general circulation, in which the momentum exchange between
eddies and the mean flow extends over layers with thickness of order
$10^6$~bar and greater, are not viable on energetic grounds: they
require energy sources several orders of magnitude larger than
Jupiter's \citep[e.g.,][]{SunZP93,Heimpel05,Heimpel07}. It also means
that the thin-shell approximation---in which the distance from any
point in the atmosphere to the planet's center is taken to be constant
and equal to the planetary radius---is adequate for and will be made
in the following considerations of the tropospheric eddy transport of
angular momentum (although the zonal flow can extend deeply, see
section~\ref{s:deep_flow}).

Meridional eddy transport of angular momentum is evidence of
meridional eddy propagation and irreversible dynamics
\citep[e.g.,][]{Edmon80}. In a thin shell, eddies that propagate
meridionally transport angular momentum in the opposite direction of
their propagation if the meridional gradient of absolute vorticity, or
depth-averaged potential vorticity, is positive (northward) between
their generation and dissipation (breaking) regions. In the
dissipation regions, irreversible meridional mixing of absolute
vorticity then leads to southward eddy vorticity fluxes and divergence
of meridional eddy angular momentum fluxes; the compensating
convergence of meridional eddy angular momentum fluxes occurs in the
generation regions, implying angular momentum transport from the
dissipation into the generation regions
(\citealp{Kuo51,Held75,Andrews76,Andrews78c,Rhines94};
\citealp[][chapter~12]{Vallis06}).  In Jupiter's upper troposphere,
the meridional gradient of absolute vorticity (and potential
vorticity) is generally positive, except in narrow latitude bands at
the centers of retrograde jets \citep{Ingersoll81,Read06}. The
direction of the observed eddy angular momentum fluxes hence indicates
generation of eddies in prograde jets, dissipation in retrograde jets,
and propagation in between.

In prograde off-equatorial jets, eddies can be generated by baroclinic
instability. Prograde jets are baroclinically more unstable than
retrograde jets if the speed (absolute velocity value) of the zonal
flow in prograde jets decreases with depth and in retrograde jets
either also decreases with depth or, with weaker vertical shear,
increases with depth. By thermal wind balance, meridional temperature
gradients along isobars then are equatorward in prograde jets and
either are also equatorward but weaker or are poleward in retrograde
jets. Indeed, in Jupiter's upper troposphere (at $\about 0.25$~bar),
meridional temperature gradients have been observed to be equatorward
in prograde off-equatorial jets and either equatorward but weaker or
poleward in retrograde off-equatorial jets [compare Fig.~6 in
\citet{Simon-Miller06} with Fig.~3 in \citet{Vasavada05}]. Thus,
meridional propagation of eddies generated by baroclinic instability
preferentially in prograde jets can account for the angular momentum
transport from retrograde into prograde off-equatorial jets, as
demonstrated in baroclinic GCMs that produce multiple jets
\citep[e.g.,][]{Williams79,OGorman08a}.

In the prograde equatorial jet, eddies can be generated by intrinsic
convective heat fluxes. Unlike in higher latitudes, Coriolis forces
are small in the equatorial region, where the Rossby number is order
one or greater. Horizontal pressure gradients there are limited by the
necessity to be balanced primarily by inertial accelerations, rather
than or in addition to Coriolis accelerations. Hence, they are small
(of order Froude number), and so are horizontal temperature gradients
on scales large enough that hydrostatic balance holds
\citep{Charney63}.\footnote{\label{fn:wtg_scaling}%
  The angular momentum and hydrostatic equations imply that, near the
  equator, horizontal variations in pressure $p$ and potential
  temperature $\theta$ scale as $\delta p/p \sim \delta\theta/\theta
  \sim \mathrm{Fr}$, where $\mathrm{Fr} = U V/(gH)$ is a Froude
  number, $H$ is the scale height, $U$ is a mean zonal velocity scale,
  and $V$ is the greater of an eddy velocity scale or mean meridional
  velocity scale \citep[cf.][]{Charney63}. For Jupiter parameters and
  with $H \sim 20\,\mathrm{km}$, $U \sim 100\,\mathrm{m\,s^{-1}}$, and
  $V \sim 10\,\mathrm{m\,s^{-1}}$ \citep{Porco03,Salyk06}, one obtains
  $\mathrm{Fr} \sim 10^{-3}$. This scaling holds where the Rossby
  number $\mathrm{Ro} = U/|fL|$, with length scale of flow variations
  $L$, is order one or greater. With $f = \beta y$ and $L \lesssim
  |y|$, it follows that the scaling holds at least within meridional
  distances $|y| \sim (U/\beta)^{1/2}$ of the equator \citep{Sobel01},
  which, for Jupiter, is within $\about 5000\,\mathrm{km}$ or $\about
  4^\circ$ of the equator.} On such large scales, therefore, diabatic
heating $Q$ cannot be balanced by temperature fluctuations or
horizontal temperature advection, as in higher latitudes. Instead, as
in the tropics of Earth's atmosphere, it is primarily balanced by the
adiabatic cooling associated with vertical motion $\omega$ acting on a
(possibly small) static stability $S$, giving the weak temperature
gradient approximation of the thermodynamic equation\footnote{%
  We use pressure coordinates, with $S = -\partial_p \theta$ and
  $\omega=Dp/Dt$. To obtain the relative magnitudes of the terms on
  the left-hand side of the thermodynamic equation $(\partial_t +
  {\vec{v} \cdot \gradh})\theta - \omega S = Q$, with isobaric
  horizontal derivative operator $\gradh$, we assume the explicit time
  derivative and horizontal advection terms scale as
  $(V/L)\delta\theta$, and the vertical advection term scales as
  $(V/L) \Delta\theta$. We have used mass conservation to relate
  vertical to horizontal velocity scales, and $\Delta\theta$ is a
  vertical potential temperature change over a scale height.  The
  vertical advection term then dominates, and the weak temperature
  gradient approximation \eqref{e:wtg} is adequate, if vertical
  potential temperature changes satisfy $\Delta\theta/\theta \gg
  \delta\theta/\theta \sim \mathrm{Fr} \sim 10^{-3}$. In Jupiter's
  equatorial troposphere, if one takes the thermal stratification
  along the Galileo probe descent path as representative, this is
  assured at least above 1.7~bar \citep{Magalhaes02}.}
\citep{Held85,Sobel01}
\begin{equation}\label{e:wtg}
  \omega S \approx -Q.
\end{equation}
By mass conservation, $\divh \vec{v}_\chi = - \partial_p \omega$,
vertical gradients of diabatic heating and/or static stability thus
induce large-scale horizontal divergence
\begin{equation}\label{e:div}
  \divh \vec{v}_\chi \approx \partial_p (Q/S),
\end{equation}
where we have decomposed the horizontal velocity $\vec{v} =
\vec{v}_\chi + \vec{v}_\Psi$ into divergent ($\vec{v}_\chi$) and
rotational ($\vec{v}_\Psi$) components. As discussed by
\citet{Sardeshmukh88}, this horizontal divergence drives a rotational
flow: Vortex stretching and vorticity advection by the divergent flow
are a source
\begin{equation}\label{e:vorticity_source}
  R = -\zeta_a (\divh \vec{v}_\chi) - (\vec{v}_\chi \cdot \gradh) \zeta_a 
\end{equation}
of rotational flow, as can be seen from the equation for the absolute
vorticity $\zeta_a = f + {\vec{k} \cdot (\curl \vec{v}_\Psi)} = f +
\zeta$ in the equatorial troposphere in the form
\begin{equation}\label{e:vorticity}
  (\partial_t + \vec{v}_\Psi \cdot \gradh) \zeta_a \approx R.
\end{equation}
Neglected here are the baroclinic term, consistent with the weak
temperature gradient approximation \citep{Charney63,Sobel01}, as well
as friction and the vertical advection and tilting terms. It follows
that convective heating fluctuations that cannot be balanced by slow
radiative processes induce fluctuations in the large-scale horizontal
divergence, and these represent a source
\begin{equation}\label{e:rossby_source}
  R' = R - \Bar R = -\divh (\zeta_a \vec{v}_\chi - \overline{\zeta_a \vec{v}_\chi})  
\end{equation}
of vorticity fluctuations and thus a source of Rossby waves
\citep{Sardeshmukh88}. (Overbars denote isobaric zonal and temporal
means and primes deviations therefrom.)

Horizontally, the Rossby wave source \eqref{e:rossby_source} can be
expected to be large in the equatorial region because the Rossby
number there is order one or greater, and large-scale horizontal flow
fluctuations induced by convective heating fluctuations are divergent
at leading order. In contrast, the Rossby number in higher latitudes
is small, large-scale horizontal flow fluctuations are nondivergent at
leading order, and baroclinic generation of vorticity fluctuations,
neglected in the Rossby wave source \eqref{e:rossby_source}, can be
important. Scale analysis\footnote{\label{fn:rossby_source}%
  For the scale analysis, we use the eddy velocity scale $V \sim
  10\,\mathrm{m\,s^{-1}}$ and the length scale of flow variations $L
  \sim 5000\,\mathrm{km}$ both for eddies and for mean fields (cf.\
  footnote~\ref{fn:wtg_scaling}); we take the scales to be invariant
  with latitude, as indicated, for the velocity scale, by the Cassini
  image analysis of \citet{Salyk06}. Then, where $\mathrm{Ro} \gtrsim
  1$ and if horizontal velocity fluctuations are divergent at leading
  order, the Rossby wave source owing to advection of planetary
  vorticity by the divergent flow is of order $R' \sim \beta V \sim 5
  \times 10^{-11}\,\mathrm{s^{-2}}$; the Rossby wave source owing to
  stretching of planetary vorticity, evaluated at $4^\circ$ latitude,
  is of the same order $R' \sim |f V|/L \sim 5 \times 10^{-11}
  \,\mathrm{s^{-2}}$. Where $\mathrm{Ro} < 1$, the divergent velocity
  is of order $\mathrm{Ro} \, V$, and the Rossby wave source is of
  order $R' \sim \mathrm{Ro} \, |fV|/L = UV/L^2$. This is of the same
  order ($R' \sim 5 \times 10^{-11}\,\mathrm{s^{-2}}$) as the Rossby
  wave source near the equator where the mean zonal velocity scale is
  of order $U \sim 100\,\mathrm{m\,s^{-1}}$ (throughout the prograde
  equatorial jet and in the strong prograde jet at $21^\circ$N, see
  Fig.~\ref{f:winds}a). At other latitudes, the mean zonal velocity
  scale ($U \lesssim 30\,\mathrm{m\,s^{-1}}$) and the Rossby wave
  source ($R' \lesssim 10^{-11}\,\mathrm{s^{-2}}$) are smaller, albeit
  only by $O(1)$ factors.%
} %
suggests that the Rossby wave source \eqref{e:rossby_source} is
largest in the prograde equatorial jet and in the strong prograde jet
at $21^\circ$N---that is, not only in the latitude band in which the
Rossby number is order one or greater (within $\about 4^\circ$ of the
equator, see footnote~\ref{fn:wtg_scaling}), but in a slightly wider
latitude band around the equator and in strong jets
generally. Vertically, the Rossby wave source \eqref{e:rossby_source}
should be largest in the upper troposphere, below the top of the
convective outflows, where fluctuations in the large-scale horizontal
divergence \eqref{e:div} can be expected to be largest: Convective
heating fluctuations and their vertical gradients can be expected to
be largest there, and the static stability has substantial vertical
gradients \citep{Magalhaes02}, marking the transition from lower
layers that are neutrally stratified by convection to upper layers
that are more stably stratified because the stabilizing radiative
heating from above begins to have an effect. The Rossby wave source
\eqref{e:rossby_source}, then, should be largest in Jupiter's
equatorial upper troposphere if convective heating fluctuations there
are sufficiently strong.

\section{Generation of equatorial
  superrotation}\label{s:superrotation}

By Hide's theorem, the eddy transport of angular momentum into the
equatorial region observed in Jupiter's upper troposphere is necessary
for the existence of a prograde equatorial jet
\citep{Hide69,SchneiderEK77b,Saravanan93,Held99c,Schneider06b}. To see
how such angular momentum transport can come about, consider the eddy
enstrophy equation implied by the vorticity equation
\eqref{e:vorticity},
\begin{equation}\label{e:enstrophy}
  \partial_t \overline{\zeta'^2}/2 + \overline{v_\Psi' \zeta'} \,
  \partial_y \Bar \zeta_a
  \approx \overline{R'\zeta'},
\end{equation}
where $y=a \phi$ is the meridional coordinate with planetary radius
$a$ and latitude $\phi$.  Neglected here is the advection of eddy
enstrophy by the eddies themselves, a triple correlation term that
seems to be about an order of magnitude smaller in Jupiter's
equatorial upper troposphere than the retained term $\overline{v_\Psi'
  \zeta'} \, \partial_y \Bar \zeta_a$.\footnote{To estimate the
  relative magnitude of the terms, we roughly approximate the absolute
  vorticity gradient $\partial_y \Bar\zeta_a$ by $\beta$
  \citep{Read06} and use the horizontal length scale of flow
  variations $L \sim 5000\,\mathrm{km}$ (see
  footnote~\ref{fn:rossby_source}) and the rotational meridional eddy
  velocity scale $V \sim 10\,\mathrm{m\,s^{-1}}$ \citep{Salyk06};
  that is, we assume the rotational and divergent velocities are of
  the same order (cf.\ footnote~\ref{fn:rossby_source}). Then, the
  magnitude of the neglected eddy enstrophy advection $\divy
  (\overline{v_\Psi' \zeta'^2})/2$ relative to the retained term
  $\overline{v_\Psi' \zeta'} \, \partial_y \Bar \zeta_a$ is of order
  $V/(\beta L^2) \sim 0.1$.} In a statistically steady state, upon
division by $\partial_{y} \Bar\zeta_a \ne 0$ and multiplication by the
thin-shell approximation of the distance $r_\perp = a \cos\phi$ to the
planet's spin axis, the eddy enstrophy equation becomes the wave
activity balance \citep{Andrews76,Andrews78c,Edmon80}
\begin{equation}\label{e:wa_balance}
  G \approx
  \overline{v_\Psi' \zeta'} \, r_\perp = - \divy(\overline{u_\Psi'
    v_\Psi'} \, r_\perp).
\end{equation}
Here, $\divy(\cdot)$ is the isobaric meridional divergence operator,
and
\begin{equation}\label{e:wa_gen}
  G = \frac{\overline{R'\zeta'}}{\partial_y \Bar\zeta_a} \, r_\perp
\end{equation}
represents the generation of wave activity $A$, with 
\begin{equation}\label{e:wa}
  A = \frac{1}{2} \, \frac{\overline{\zeta'^2}}{\partial_y
    \Bar\zeta_a} \, r_\perp.
\end{equation}  
The wave activity balance \eqref{e:wa_balance} states that at
latitudes at which wave activity is generated ($G>0$), the eddy
vorticity flux $\overline{v_\Psi' \zeta'}$ is directed northward,
implying convergence of (rotational) eddy angular momentum fluxes
$\overline{u_\Psi' v_\Psi'} \, r_\perp$ (per unit mass); conversely,
at latitudes at which wave activity is dissipated ($G < 0$), there is
divergence of eddy angular momentum fluxes.  The meridional eddy
angular momentum flux is equal to minus the meridional wave activity
flux.
Thus, generation of wave activity in the equatorial region and
radiation to and dissipation in higher latitudes entail angular
momentum transport from higher latitudes into the equatorial
region. See \citet{Andrews76,Andrews78c}, \citet{Plumb79},
\citet{McIntyre80b}, \citet{Edmon80}, and
\citet[][chapter~7]{Vallis06} for further discussion and
generalizations of these results from wave--mean flow interaction
theory; and see \citet{Suarez92} and \citet{Saravanan93} for numerical
demonstrations that (stationary) Rossby wave sources in the equatorial
region entail convergence of eddy angular momentum fluxes and can
generate superrotation.

The wave activity balance \eqref{e:wa_balance} implies that in
Jupiter's upper troposphere, convergence of eddy angular momentum
fluxes in the equatorial region is to be expected if Rossby wave
generation by convective heating fluctuations is sufficiently
strong. Contributions to the Rossby wave source $R'$ that do not
depend on vorticity fluctuations $\zeta'$, to the extent that
$\partial_t \zeta' \sim R'$, can be expected to contribute positively
to the eddy enstrophy generation $\overline{R'\zeta'}$ and thus,
because the absolute vorticity gradient is positive \citep{Read06}, to
the wave activity generation $G$. Some of the wave activity so
generated may dissipate near or in its generation region, as
contributions to the Rossby wave source $R'$ that depend on vorticity
fluctuations $\zeta'$ may damp them. The principal damping term is
$-(\divy \Bar v_{\chi}) \overline{\zeta'^2}$; as pointed out
\citet{Sardeshmukh88}, this term contributes negatively to the eddy
enstrophy and wave activity generation where the mean meridional flow
is divergent, for example, in the horizontal outflows of ascending
branches of any (Eulerian) mean meridional circulation cells. Near the
equator, however, the damping limits itself as the strength of
convective heating fluctuations increases: Wave activity generation
near the equator entails convergence of eddy angular momentum fluxes,
which implies weakened poleward or even equatorward mean meridional
flow, as can be seen from the zonally and temporally averaged angular
momentum balance in a statistically steady state,
\begin{equation}\label{e:am_balance_loc}
  -\Bar{\vec{u}} \cdot \grad M_\Omega 
  = f \Bar v_\chi \, r_\perp  = \mathcal{S}_e + \mathcal{S}_m.
\end{equation}
Here, $\vec{u}$ is the three-dimensional velocity vector, $M_\Omega =
\Omega r_\perp^2$ is the angular momentum per unit mass owing to the
planetary rotation (with constant angular velocity $\Omega$), and the first
equality holds in the thin-shell approximation. We have again
neglected friction, we have used the fact that the mean meridional
flow is irrotational ($\Bar v = \Bar v_\chi$), and
\begin{subequations}
  \begin{equation}
    \mathcal{S}_e = \divy(\overline{u' v'} \, r_\perp)
    + \partial_p (\overline{u' \omega'} \, r_\perp),
  \end{equation}
  and
  \begin{equation}
    \mathcal{S}_m = \divy(\Bar{u} \, \Bar{v} \, r_\perp)
    + \partial_p (\Bar u \, \Bar \omega \, r_\perp),
  \end{equation}
\end{subequations}
are the divergences of fluxes of relative angular momentum per unit
mass $M_u = u r_\perp$ owing to eddies and mean meridional
circulations.  The angular momentum balance \eqref{e:am_balance_loc}
shows that as the strength of convective heating fluctuations
increases, increasing wave activity generation and the increasing eddy
angular momentum flux convergence (decreasing $\mathcal{S}_e$) it
entails either weaken any divergence of the mean meridional flow at
the equator or even lead to convergence and thus to eddy enstrophy and
wave activity amplification.\footnote{This is one of several eddy-mean
  flow feedbacks that can lead to rapid transitions to superrotation
  as parameters such as the strength of the convective heating are
  varied. Other feedbacks are discussed by \citet{Saravanan93} and
  \citet{Held99c}.}  Sufficiently strong convective heating
fluctuations, then, can be expected to lead to net wave activity
generation ($G > 0$) and convergence of eddy angular momentum fluxes
in the equatorial region.

Convergence of eddy angular momentum fluxes in the equatorial region
accelerates the zonal flow. As a consequence, a prograde equatorial
jet forms if other processes that may decelerate it---e.g., drag at
depth linked to upper-tropospheric dynamics through mean meridional
circulations (see section~\ref{s:deep_flow})---are sufficiently
weak. Such an equatorial jet can be expected to occupy a latitude band
at least as wide as the meridional decay scale of long equatorial
Rossby waves: the equatorial Rossby radius $L_e = (c/\beta)^{1/2}$
with tropospheric gravity wave speed $c$ (\citealp{Matsuno66};
\citealp[chapter~11]{Gill82}). For Jupiter, if one takes the speed $c
\approx 450\,\mathrm{m\,s^{-1}}$ of the waves observed after the
impact of comet Shoemaker-Levy~9 as the relevant gravity wave speed
\citep{Ingersoll95}---a wave speed roughly consistent with the
tropospheric thermal stratification along the Galileo probe descent
path \citep[cf.][]{Magalhaes02}---one obtains $L_e \approx
9500\,\mathrm{km}$, or a lower bound on the half-width of the jet of
$\about 8^\circ$. This is similar to, albeit smaller than, the
half-width of Jupiter's equatorial jet in the upper troposphere
(Fig.~\ref{f:winds}a).\footnote{\label{fn:width_scales}%
  Other lower bounds on the half-width of the equatorial jet are the
  half-widths of the latitude bands with (i) substantial fluctuations
  in the horizontal divergence or (ii) relative angular momentum
  transport by mean meridional circulations. Both latitude bands are
  characterized by $\mathrm{Ro} \gtrsim 1$
  \citep[e.g.,][]{Sobel01,Schneider06b}. With the upper bound $U
  \lesssim c/2$ on the equatorial jet speed given below, one finds
  that the half-width of the latitude band with $\mathrm{Ro} \gtrsim
  1$ is constrained to be of the same order of but smaller than $L_e$
  (cf.\ footnote~\ref{fn:wtg_scaling}).}

An upper bound on the speed of the equatorial jet can be obtained by
vorticity homogenization arguments. The jet speed is determined by the
change in vorticity within the equatorial jet owing to the meridional
redistribution of absolute vorticity relative to a state of solid body
rotation. If the zonal flow is zero at a distance $L_s$ from the
equator and if this distance is small enough that the small-angle
approximation for latitudes is adequate, the jet speed at the equator
is $U \approx \Delta\zeta L_s$, where $\Delta\zeta$ is the absolute
value of the relative vorticity averaged between the equator and a
distance $L_s$ away from it. The absolute vorticity redistribution
reaches an end state, and meridional Rossby wave propagation ceases,
when the absolute vorticity within the jet is homogenized in each
hemisphere. If the absolute vorticity at the zeros of the jet is
approximated by $\pm \beta L_s$, the state of homogenized absolute
vorticity corresponding to maximum prograde jet speed for a given jet
width has absolute vorticity $+\beta L_s$ in the northern and $-\beta
L_s$ in the southern hemisphere, with a barotropically stable jump at
the equator: an absolute vorticity ``staircase''
\citep{McIntyre82,Dritschel08}. The absolute value of the relative
vorticity averaged between the equator and a distance $L_s$ away from
it then is $\Delta\zeta \approx \beta L_s/2$, and one obtains the
upper bound on the jet speed
\begin{equation}\label{e:jet_speed}
  U \lesssim \frac{\beta L_s^2}{2} .
\end{equation}
If one further substitutes $L_s \sim L_e$, one obtains $U \lesssim
c/2$, that is, a bound on the jet speed that only depends on the
gravity wave speed. For Jupiter, with $c \approx
450\,\mathrm{m\,s^{-1}}$, this gives $U \lesssim
225\,\mathrm{m\,s^{-1}}$, which is of the same order of, albeit larger
than, the observed equatorial jet speed in the upper troposphere
(Fig.~\ref{f:winds}a); it is closer to the equatorial jet speed at
greater depth along the Galileo probe descent path. If the actual
half-width of Jupiter's equatorial jet is used ($L_s \gtrsim L_e$),
the overestimation of the jet speed in the upper troposphere is more
pronounced; a state of homogenized absolute vorticity in each
hemisphere is not attained in the upper troposphere
\citep{Read06}. Nonetheless, these arguments show that the speed of
prograde equatorial jets, to the extent that a state of homogenized
absolute vorticity in each hemisphere is being approached, can be
expected to increase roughly linearly with $\beta$ and quadratically
with their half-width. The speed of such jets does not depend
directly, as is sometimes surmised, on the total energy uptake by the
atmosphere or the total kinetic energy dissipation, but on emergent
properties such as the thermal stratification, which determines the
gravity wave speed. Consistent with these arguments and the similar
radii and rotation rates of Jupiter and Saturn, Saturn's prograde
equatorial jet is about 2 times wider and 3--4 times stronger than
Jupiter's. This may be a consequence of a greater gravity wave speed
on Saturn.

\section{Drag at depth and mean meridional
  circulations}\label{s:deep_flow}

In a statistically steady state, eddy angular momentum transport from
retrograde into prograde jets in the upper troposphere must be
balanced by angular momentum transport by mean meridional circulations
or eddies in other layers or by drag on the zonal flow. Drag acts on
the zonal flow deep in Jupiter's atmosphere, where hydrogen, its
primary constituent, undergoes a continuous transition from an
electrically semi-conducting molecular phase in the outer layers to a
conducting metallic phase in the interior
\citep{Guillot04,Guillot05}. Where the atmosphere is electrically
conducting, any flow advects the magnetic field and induces an
electric current. The Ohmic dissipation of the induced current implies
a magnetohydrodynamic (MHD) drag on the flow
\citep{Grote01,Liu08}. This MHD drag we take to provide the principal
angular momentum dissipation mechanism on Jupiter. Because the
electric conductivity increases continuously with depth, it is
difficult to estimate where in the atmosphere the MHD drag acts. For
the sake of argument here, we assume that substantial MHD drag is
confined within $\about 0.96$ Jupiter radii (pressures greater than
$\about 10^5$~bar). This corresponds to the estimated maximum depth up
to which zonal flows with speeds similar to those in the upper
troposphere could extend before the energy dissipation by the MHD drag
would violate the constraint that it cannot exceed Jupiter's
luminosity \citep{Liu08}. Precisely at which depth the MHD drag acts
is not essential for our arguments, but it is crucial that there is an
outer atmospheric shell in which the flow is effectively frictionless.

Because the MHD drag acts at great depth, deep-atmosphere dynamics
must be considered in linking it to dynamics in the upper
troposphere. It is well known that in a thin atmospheric shell, in the
zonal, temporal, and vertical mean in a statistically steady state,
drag on the zonal flow balances any transport of angular momentum into
or out of a latitude band \citep[e.g.,][chapter~11]{Peixoto92}; that
is, the zonally, temporally, and vertically averaged angular momentum
balance is
\begin{equation}\label{e:am_balance_vert}
  \< \mathcal{S}_e + \mathcal{S}_m \> = r_\perp \< \Bar{\mathcal{D}} \>,
\end{equation}
where $\< \cdot \>$ denotes a mass-weighted vertical mean and
$\mathcal{D}$ is the drag force per unit mass on the zonal flow, for
example, $\mathcal{D} = - k u$ with relaxation coefficient $k$ for
Rayleigh drag. This generalizes to a deep atmosphere if $r_\perp = r
\cos\phi$ is taken to be the actual distance to the planet's spin
axis, with distance to the planet's center $r$, and if the vertical
mean is understood as a mean along surfaces of constant planetary
angular momentum per unit mass $M_\Omega = \Omega r_\perp^2$. Such
angular momentum surfaces are vertical in the thin-shell approximation
($r=a=\mathrm{const}$) but are parallel to the planet's spin axis in a
deep atmosphere. That is, the zonal and vertical mean in a thin
atmosphere must be replaced by a mean along cylinders concentric with
the planet's spin axis in a deep atmosphere (see appendix~A).

\begin{figure}[!tb]
  \centering\includegraphics{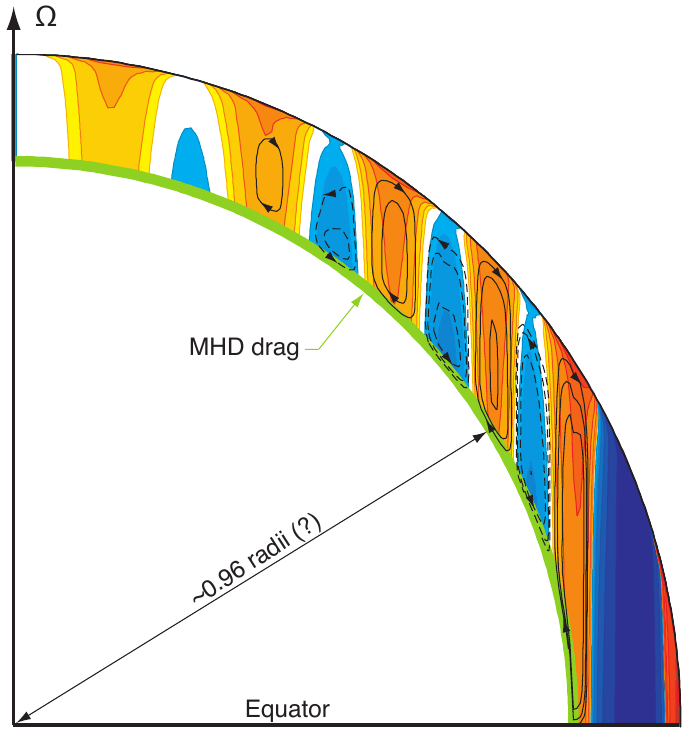}
  \caption{Schematic of Jupiter's envisaged zonal flow and mean
    meridional circulations. Shown is one quadrant in the meridional
    plane (not to scale). Colors indicate the zonal flow (yellow/red
    for prograde and cyan/blue for retrograde flow). Contours indicate
    the mass flux streamfunction of the mean meridional circulation
    (solid for clockwise rotation and dashed for counterclockwise
    rotation). We have omitted mean meridional circulations in the
    equatorial region in which cylinders concentric with the planet's
    spin axis do not intersect the layer with MHD drag; they likely
    have a more complex structure than those in higher latitudes.}
  \label{f:sketch}
\end{figure}
In the equatorial region in which cylinders concentric with the
planet's spin axis do not intersect the layer with MHD drag, there is
effectively (absent other dissipation mechanisms except weak viscous
dissipation) no cylindrically averaged drag $\<\Bar{\mathcal{D}}\>$ on
the zonal flow. If substantial MHD drag is confined within $\about
0.96$ Jupiter radii, the region of vanishing cylindrically averaged
MHD drag extends in the outer atmosphere from the equator to $\about
16^\circ$ latitude in each hemisphere and projects from there downward
along cylinders (see the schematic in Fig.~\ref{f:sketch}). In this
region, the angular momentum balance \eqref{e:am_balance_vert} implies
that any convergence of eddy angular momentum fluxes in the upper
troposphere can only be balanced by divergence of eddy angular
momentum fluxes in other layers and/or by divergence of relative
angular momentum fluxes owing to mean meridional circulations. The
mean meridional circulations likely cannot link equatorial convergence
of eddy angular momentum fluxes in the upper troposphere to the MHD
drag at depth, as considerations of Rossby numbers show. Mean flows in
the meridional direction can only lead to substantial divergence of
relative angular momentum fluxes where the Rossby number $\mathrm{Ro}
= U/|fL|$, with meridional length scale of zonal-flow variations $L$,
is order one or greater. The region where $\mathrm{Ro} \gtrsim 1$ has
meridional half-width $\lesssim L_e$ (footnote~\ref{fn:width_scales}),
even at depths at which the zonal flow speed may exceed its
upper-tropospheric value but remains constrained by the bound
\eqref{e:jet_speed}.  It extends to $\lesssim 8^\circ$ latitude in the
outer atmosphere, so the latitude band in which $\mathrm{Ro} \gtrsim
1$ is narrower than the latitude band of vanishing cylindrically
averaged MHD drag provided there is no MHD drag in an outer
atmospheric shell at least $\about 0.01$ Jupiter radii
thick. Similarly, mean flows in the cylindrically-radial direction
(perpendicular to the planet's spin axis) can only lead to substantial
divergence of relative angular momentum fluxes where the Rossby number
$\mathrm{Ro}_\perp = U/(2\Omega L_\perp)$, with cylindrically-radial
length scale of zonal-flow variations $L_\perp$, is order one or
greater (see appendix~A). With velocity scale $U \sim
200\,\mathrm{m\,s^{-1}}$ and length scale $L_\perp \sim
2000\,\mathrm{km}$, corresponding to the radial length scale in the
equatorial plane of a jet that extends from the equator to $\about
13^\circ$ latitude in the outer atmosphere and projects downward along
cylinders, this Rossby number is $\mathrm{Ro}_\perp \sim 0.3$; it is
even smaller in higher latitudes where the zonal-flow velocity must be
smaller because it cannot substantially exceed its upper-tropospheric
values of order $10\,\mathrm{m\,s^{-1}}$ without violating the
constraint that the energy dissipation by the MHD drag cannot exceed
Jupiter's luminosity \citep{Liu08}. Therefore, mean meridional
circulations can redistribute angular momentum only within a region
that is narrower than that in which cylinders concentric with the
planet's spin axis do not intersect the layer with MHD drag, provided
there is no MHD drag in an outer atmospheric shell at least $\about
0.01$ Jupiter radii thick. So they likely cannot link equatorial
convergence of eddy angular momentum fluxes resulting from meridional
radiation of convectively generated Rossby waves to the MHD drag at
depth. To the extent that divergence of eddy angular momentum fluxes
outside the layer of convective Rossby wave generation and divergence
of relative angular momentum fluxes owing to mean meridional
circulations merely compensate the acceleration of the cylindrically
averaged zonal flow, generation of equatorial superrotation by Rossby
wave radiation seems inevitable.

In higher latitudes, where cylinders concentric with the planet's spin
axis do intersect the layer with MHD drag, mean meridional
circulations link dynamics in the upper troposphere to the MHD drag at
depth. The Rossby numbers $\mathrm{Ro}$ and $\mathrm{Ro}_\perp$ in
higher latitudes are small, both in the upper troposphere and in
deeper layers. Therefore, the relative angular momentum flux
divergence owing to eddies dominates that owing to mean meridional
circulations, and the local angular momentum balance
\eqref{e:am_balance_loc} in the upper troposphere, where there is
effectively no drag on the zonal flow, reduces to
\begin{equation}\label{e:low_Rossby_am}
  f \Bar v_\chi \, r_\perp  \approx \mathcal{S}_e. 
\end{equation}
It follows that in regions of eddy angular momentum flux convergence
(prograde jets), the mean meridional mass flux is equatorward; in
regions of divergence (retrograde jets), it is poleward. Between the
layer with substantial eddy angular momentum fluxes and the layer with
MHD drag, provided the divergence of convective angular momentum
fluxes (Reynolds stress) is negligible, the mean meridional
circulations are unaffected by zonal torques ($S_e \approx r_\perp
\mathcal{D} \approx 0$); the local angular momentum balance
\eqref{e:am_balance_loc} in a form suitable for a deep atmosphere can
then be expressed as
\begin{equation}
  \Bar{\vec{u}} \cdot \grad \Bar M \approx 0,
\end{equation}
where $M = M_\Omega + M_u$ is the angular momentum per unit mass. It
follows that the mean meridional circulations extend downward along
surfaces of constant angular momentum per unit mass $M$
\citep[e.g.,][]{Haynes91}. 
Approximately, these angular momentum surfaces are again cylinders
concentric with the spin axis because small Rossby numbers mean that
the angular momentum $M$ is dominated by its planetary component
$M_\Omega$. Irrespective of the depth at which the MHD drag acts, the
mean meridional circulations must extend downward to and must close
where the drag allows mass fluxes to cross angular momentum surfaces
\citep{Haynes91,OGorman08a}. There, the Coriolis torque on any mass
flux component normal to angular momentum surfaces is balanced by the
MHD drag on the zonal flow (Ekman balance),
\begin{equation}\label{e:Ekman}
  \Bar{\vec{u}} \cdot \grad M_\Omega \approx r_\perp \mathcal{D}.  
\end{equation}
Figure~\ref{f:sketch} sketches the resulting mean meridional
circulations.

Ekman balance \eqref{e:Ekman} at depth requires that the zonal flow is
prograde ($\mathcal{D} < 0$) where the mean meridional mass flux has a
component toward the spin axis and retrograde ($\mathcal{D} > 0$)
where it has a component away from the spin axis. Given the
correlation between the zonal velocity and the convergence of eddy
angular momentum fluxes in the upper branches of the mean meridional
circulations, this implies that the signs and zeros of the zonal flow,
like the mean meridional circulations, project downward approximately
along cylinders, as sketched in Fig.~\ref{f:sketch}. Thermal wind
balance then constrains the thermal structure of the atmosphere below
the layer with substantial eddy angular momentum fluxes [see, e.g.,
\citet{Ingersoll82} and \citet{Kaspi08} for thermal wind equations for
deep atmospheres]. The mean meridional circulations adjust entropy
gradients and the zonal flow in lower layers such that they satisfy,
in a statistically steady state, the constraints that (i) angular
momentum flux convergence or divergence and the MHD drag on the zonal
flow balance upon averaging over cylinders, and (ii) the zonal flow is
in thermal wind balance with the entropy gradients \citep[see,
e.g.,][]{Haynes91}. These dual constraints generally cannot be
satisfied, as is often assumed in Jupiter models, with entropy
gradients that vanish throughout the deep atmosphere and with a
corresponding zonal flow without shear in the direction of the spin
axis (Taylor columns). The zonal flow speed within $\about 0.96$
Jupiter radii is constrained to be smaller than than that in the
prograde off-equatorial jets in the upper troposphere because
otherwise the energy dissipation by the MHD drag would exceed
Jupiter's luminosity \citep{Liu08}. Therefore, the zonal flow
generally must be sheared and must be associated with nonzero entropy
gradients. For example, it is conceivable that convection deep in
Jupiter's atmosphere homogenizes entropy along angular momentum
surfaces but that there are nonzero (albeit possibly weak) entropy
gradients normal to them, such that the mean thermal structure at
depth is neutral with respect to slantwise convective, or symmetric,
instability.

\section{General circulation model}\label{s:gcm}

To demonstrate the viability of the mechanisms proposed, we
constructed a GCM of Jupiter's outer atmosphere. Simulating Jupiter's
deep atmosphere in a manner that is consistent with its measured
energy balance is computationally prohibitive. Instead, our GCM is
based on the hydrostatic primitive equations for a dry ideal-gas
atmosphere in a thin spherical shell with Jupiter's radius, rotation
rate, gravitational acceleration, and thermal properties. The shell
extends from the top of the atmosphere to a lower boundary with mean
pressure 3~bar. Within this pressure range, about 90\% of the solar
radiation incident at the top of Jupiter's atmosphere is absorbed or
scattered back, and latent heat release in phase changes of water,
ammonia, and hydrogen sulfide is negligible: moist adiabatic and dry
adiabatic temperature lapse rates are nearly indistinguishable
\citep{Showman98,Ingersoll04}. Nonetheless, the idealization of
focusing on a thin shell in Jupiter's outer atmosphere means that we
are not able to resolve details of the coupling between the flow in
the outer atmosphere and that at depth, beyond the previously
discussed constraints on this coupling implied by the angular momentum
balance.

The GCM uses the spectral-transform method in the horizontal
(resolution T213) and finite differences in the vertical (30
levels). Radiative transfer is represented as that in a homogeneous
gray atmosphere, with absorption and scattering of solar radiation and
absorption and emission of thermal radiation. Optical properties of
the atmosphere are so specified that the idealized representation of
radiative transfer is qualitatively consistent with measured radiative
fluxes inside Jupiter's atmosphere \citep{Sromovsky98} and
quantitatively consistent with the measured energy balance at the top
of the atmosphere \citep{Hanel81}. At the GCM's lower boundary, a
spatially uniform and temporally constant intrinsic heat flux is
imposed. If the intrinsic heat flux is sufficiently strong to
destabilize the atmospheric thermal stratification, the convection
that ensues is represented by a quasi-equilibrium convection scheme
that relaxes temperatures in statically unstable parts of atmospheric
columns to a profile with neutral static stability
\citep{Schneider06a}. Near the lower boundary, Rayleigh drag in the
horizontal momentum equations is chosen as an idealized linear
representation of effects of the MHD drag deep in Jupiter's atmosphere
on the flow in the outer atmosphere. To represent the downward
projection along cylinders in a deep atmosphere in the thin-shell
approximation with vertical planetary angular momentum surfaces, we
use a Rayleigh drag coefficient that is constant along the GCM's lower
boundary poleward of $16.3^\circ$ latitude and rapidly decreases to
zero at lower latitudes, corresponding to the assumption that
substantial MHD drag is confined within $0.96$ Jupiter radii. Above
the layer with Rayleigh drag, highly scale-selective horizontal
hyperdiffusion at small scales \citep{Smith02b}, representing
subgrid-scale processes, is the only frictional process; in
particular, there is no vertical viscous transfer of momentum or
heat. All forcings and boundary conditions are temporally constant and
zonally and hemispherically symmetric, so the simulations have
stationary and zonally and hemispherically symmetric flow
statistics. See appendix~B for details of the GCM and simulations.

\section{Jupiter simulation}\label{s:jupiter_sim}

\subsection{Upper-tropospheric dynamics}

A simulation with Jupiter's solar constant, with insolation at the top
of the atmosphere varying with the cosine of latitude, and with an
intrinsic heat flux of $5.7\,\mathrm{W\,m^{-2}}$ \citep{Gierasch00} at
the lower boundary reproduces the large-scale features of Jupiter's
upper-tropospheric jets. It produces a prograde equatorial jet and an
alternating sequence of retrograde and prograde off-equatorial jets,
with speeds and widths similar to those on Jupiter
(Fig.~\ref{f:winds}a). The jets maintain their speeds and structure
over time in the statistically steady state of the simulation;
however, there is some low-frequency variability, most pronounced in
low latitudes, with decadal and possibly longer timescales. The
equatorial jet is wider and stronger than the off-equatorial jets and
resembles Jupiter's equatorial jet, though it does not exhibit
Jupiter's local velocity minimum at the equator. As on Jupiter,
retrograde off-equatorial jets are weaker than prograde jets, except
for the first retrograde jets off the equator; in high latitudes, some
retrograde jets are more manifest as local zonal velocity minima than
as actual retrograde flow in the upper troposphere. (The retrograde
jets are stronger at lower levels, but they are still weaker than the
prograde jets within our GCM domain; see
Fig.~\ref{f:vertical_structure} below.)  Prograde off-equatorial jets
are sharper than retrograde jets, consistent with them being
barotropically more stable \citep{Rhines94}. The meridional gradient
of absolute vorticity in the upper troposphere is generally positive
but, particularly in shorter-term averages, reverses in the sharpest
retrograde jets, where it can approach $-\beta/2$---similar to, but of
smaller magnitude than, the reversed absolute vorticity gradients in
Jupiter's retrograde jets \citep{Ingersoll81,Read06}. That Jupiter's
retrograde jets violate sufficient conditions for linear barotropic
stability in the upper troposphere has been much discussed and is
sometimes deemed puzzling
\citep[e.g.,][]{Ingersoll04,Vasavada05}. However, as demonstrated here
and previously, for example, by \citet{Williams02} and \citet{Kaspi07}
for baroclinic flows and by \citet{Marston08} for barotropic flows, it
is not necessary that statistically steady states of
forced-dissipative flows satisfy barotropic stability conditions for
unforced and non-dissipative flows.

The speeds and widths of the off-equatorial jets increase with
decreasing Rayleigh drag coefficient, similar to the drag dependence
of two-dimensional turbulent flows described by \citet{Smith02b} and
\citet{Danilov02}. Because the drag coefficient is poorly constrained
by data, we empirically chose a value ($0.05\,\mathrm{day^{-1}}$ with
$1\,\mathrm{day}=86400\,\mathrm{s}$) that resulted in a good fit to
Jupiter's off-equatorial jet speeds and widths. We also experimented
with different widths of the region of vanishing Rayleigh
drag. Doubling the width of this region such that it extends to
$33^\circ$ latitude in each hemisphere leads to a prograde equatorial
jet that is $\about 35\,\mathrm{m\,s^{-1}}$ stronger at the equator
and whose first zero off the equator is $\about 5^\circ$ farther
poleward, giving a closer match to Jupiter's equatorial jet; however,
it also leads to the first retrograde jets off the equator being
considerably stronger and wider than those on Jupiter. Nonetheless, it
is clear that the width of the region of vanishing drag does not alone
control the width of the prograde equatorial jet. But having a
sufficiently wide region of vanishing or very weak drag around the
equator is necessary to obtain equatorial superrotation with intrinsic
heat fluxes comparable to Jupiter's; if the drag coefficient in the
equatorial region is similar to that in higher latitudes, considerably
stronger intrinsic heat fluxes are necessary to generate equatorial
superrotation (see appendix~B for details).

\begin{figure*}[!tbh]
  \centering\resizebox{16.5cm}{!}{\includegraphics{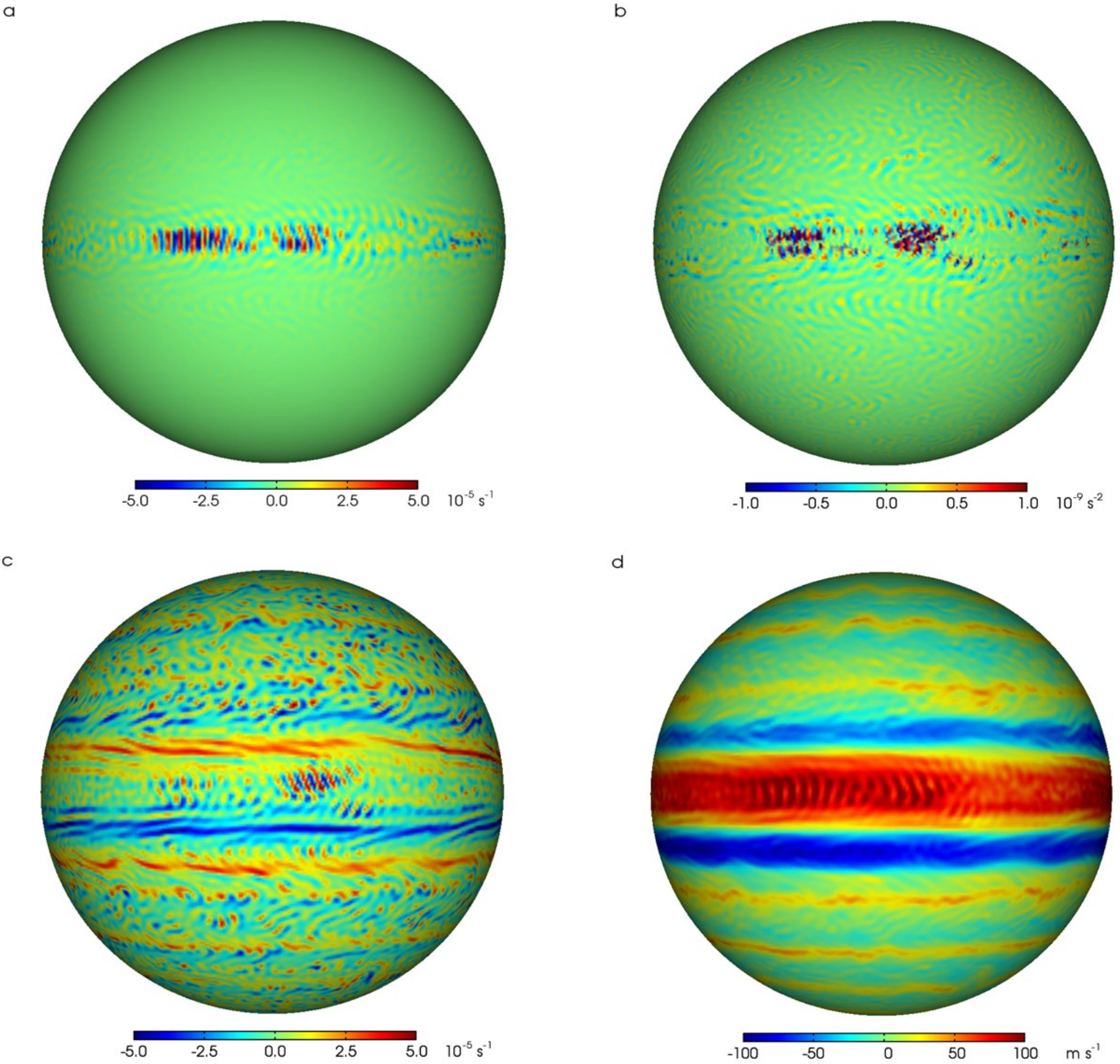}}
  \caption{Flow fields at 0.65~bar at one instant in Jupiter
    simulation. (a) Horizontal divergence. (b) Rossby wave source
    \eqref{e:rossby_source}. (c) Relative vorticity of horizontal
    flow. (d) Zonal velocity. The instant shown is within the period
    for which the mean zonal flow is shown in Fig.~\ref{f:winds}a.  }
  \label{f:snapshots}
\end{figure*}

Instantaneous flow fields in the upper troposphere provide the first
evidence that the equatorial superrotation is indeed a consequence of
meridional radiation of convectively generated Rossby waves. Although
the imposed intrinsic heat flux at the lower boundary is spatially
uniform, fluctuations in the horizontal divergence associated with
convective heating fluctuations are primarily confined to an
equatorial latitude band and there are typically modulated by
large-scale waves; the horizontal flow away from the equator is
geostrophic and thus nondivergent to leading order
(Fig.~\ref{f:snapshots}a). Root-mean-square (rms) fluctuations in the
horizontal divergence are maximal at the equator, where they reach
$\about 1\times 10^{-5} \, \mathrm{s^{-1}}$; they decay to half their
maximum value at $\about 4^\circ$ latitude, consistent with our
scaling estimates (footnote~\ref{fn:wtg_scaling}). In the vertical,
rms fluctuations in the horizontal divergence are maximal in the upper
troposphere, below the top of the convective outflows (near and
slightly above the level at which they are shown in
Fig.~\ref{f:snapshots}). Similarly, the Rossby wave source
\eqref{e:rossby_source} is also primarily confined to an equatorial
latitude band (Fig.~\ref{f:snapshots}b). In the vertical, rms
fluctuations in the Rossby wave source \eqref{e:rossby_source} are
maximal in the upper troposphere, like the horizontal divergence
fluctuations. In the horizontal, they are maximal $\about 2^\circ$ off
the equator, where they can be generated by vortex stretching and
reach $\about 1.5 \times 10^{-10}\,\mathrm{s^{-2}}$; they decay to
half their maximum value at $\about 7^\circ$ latitude, consistent with
our scaling estimates (footnote~\ref{fn:rossby_source}). While
fluctuations in the horizontal divergence and in the Rossby wave
source \eqref{e:rossby_source} are primarily confined to the
equatorial region, the vorticity of the horizontal flow shows eddies
at all latitudes (Fig.~\ref{f:snapshots}c). The vorticity fluctuations
are of similar magnitude as divergence fluctuations near the equator
but are much larger than divergence fluctuations away from the
equator---consistent with eddy generation by convective heating
fluctuations near the equator and eddy generation by baroclinic
instability away from the equator. The vorticity field clearly shows
the shear zones between the zonal jets as well as smaller-scale
coherent vortices, though no large coherent vortices such as the Great
Red Spot. It is possible that large vortices such as the Great Red
Spot form spontaneously but would require longer integration times
than we can achieve in our simulation, or that deep-atmosphere
dynamics not captured in our simulation are important for their
formation and stability. That the zonal jets are present and coherent
at every instant, not only upon averaging, is most clearly evident in
the zonal velocity field, which also shows the equatorial waves
recognizable in the other flow fields, as well as undulations of
off-equatorial jets (Fig.~\ref{f:snapshots}d). Animations (available
at www.gps.caltech.edu/$\sim$tapio/papers/) show that the equatorial
waves, organized into large wave packets, exhibit retrograde phase
velocities, consistent with them being Rossby waves. The retrograde
tilt of their phase lines away from the equator
(Fig.~\ref{f:snapshots}d) indicates that they transport angular
momentum toward the equator \citep[cf.][chapter~11]{Peixoto92}.

\subsection{Vertical structure and angular momentum fluxes}

\begin{figure*}[!tb]
  \centering\includegraphics{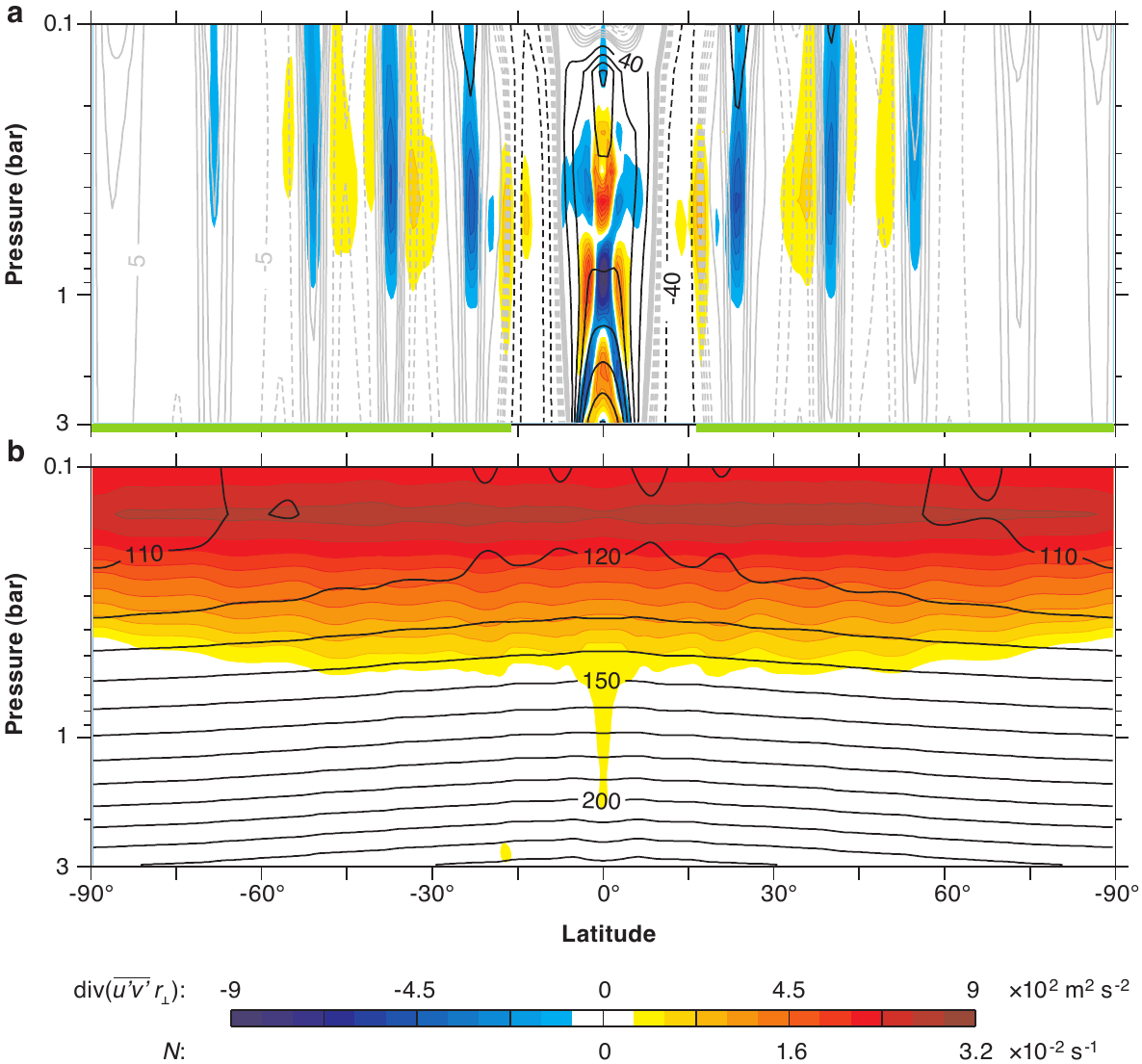}
  \caption{Flow fields in the latitude-pressure plane in Jupiter
    simulation. (a) Zonal flow (contours) and divergence $\divy
    (\overline{u'v'} \, r_\perp)$ of meridional eddy angular momentum
    fluxes (colors). Gray contours for zonal flow speeds between $5$
    and $20\,\mathrm{m\,s^{-1}}$, with a contour interval of
    $5\,\mathrm{m\,s^{-1}}$; black contours for zonal flow speeds of
    $40\,\mathrm{m\,s^{-1}}$ or above, with a contour interval of
    $20\,\mathrm{m\,s^{-1}}$. Solid contours for prograde flow and
    dashed contours for retrograde flow.  (b) Temperature (contours,
    contour interval $10\,\mathrm{K}$) and Brunt-V{\"a}is{\"a}l{\"a}
    frequency $N$ (colors). Shown in this and subsequent figures are
    zonal and temporal means over the same 1500 days for which the
    zonal flow at 0.65~bar is shown in Fig.~\ref{f:winds}a. The green
    part of the latitude axis marks the latitudes with Rayleigh
    drag. The graphs are truncated at 0.1~bar at the top, but the
    uppermost full level of the GCM has a mean pressure of 0.05~bar.}
  \label{f:vertical_structure}
\end{figure*}
The vertical structure of the zonal flow in the simulation indicates
preferential baroclinic eddy generation in prograde off-equatorial
jets and is consistent with what is known about Jupiter's equatorial
jet in lower layers (Fig.~\ref{f:vertical_structure}a). The speed of
the prograde equatorial jet increases with depth, for example, at the
equator, from $\about 80\,\mathrm{m\,s^{-1}}$ between 0.25 and 1~bar
to $\about 160\,\mathrm{m\,s^{-1}}$ at 3~bar. This is similar to the
increase in zonal flow speed seen on Jupiter along the Galileo probe
descent path at $6.4^\circ$N \citep{Atkinson98}, which lies within
Jupiter's equatorial jet and so may be comparable to latitudes closer
to the equator in our simulation with a narrower equatorial jet. The
speed of the off-equatorial zonal flow decreases with depth in
prograde jets and increases, with weaker vertical shear, in retrograde
jets. For example, the maximum prograde jet speed poleward of
$20^\circ$ latitude decreases from $\mathrm{30}\,\mathrm{m\,s^{-1}}$
at 0.65~bar to $\mathrm{23}\,\mathrm{m\,s^{-1}}$ at 3~bar, whereas the
maximum retrograde jet speed increases from
$\mathrm{13}\,\mathrm{m\,s^{-1}}$ at 0.65~bar to
$\mathrm{15}\,\mathrm{m\,s^{-1}}$ at 3~bar. This implies that the
prograde off-equatorial jets are baroclinically more unstable than the
retrograde jets.

The thermal structure of the atmosphere is in thermal wind balance
with the zonal flow, shows the signature of convection penetrating
into the upper troposphere, and is consistent with what is known about
Jupiter's thermal structure (Fig.~\ref{f:vertical_structure}b). For
example, meridional temperature gradients along isobars are poleward
(reversed) within the equatorial jet, consistent with thermal wind
balance and zonal flow speeds increasing with depth. Away from the
equator, they are equatorward in prograde jets but are poleward or
close to zero in many retrograde jets, consistent with the opposite
signs of the vertical shear in prograde and retrograde jets. The
atmosphere is stably stratified above $\about 0.5$~bar, with a
temperature inversion above $\about 0.2$~bar and with a level of
minimum temperature lapse rate (maximum Brunt-V{\"a}is{\"a}l{\"a}
frequency) at $\about 0.15$~bar that may be identified with the
tropopause. These features of the thermal structure are qualitatively
and to a large degree quantitatively consistent with the available
data for Jupiter's upper troposphere \citep{Simon-Miller06}. A
possible quantitative discrepancy may be that in the layer between
$0.1$ and $0.5$~bar, which contains the thermal emission level on
Jupiter as well as in the simulation, pole-equator temperature
contrasts along isobars may be a few Kelvin larger in the simulation
than on Jupiter. (If it does not result from data and retrieval
limitations, the lack of vertical resolution and of representation of
hazes in our GCM may account for this discrepancy.)  Beneath $\about
0.5$~bar, the atmosphere in the simulation is nearly neutrally
stratified by convection. As a consequence of the nearly neutral
stratification, meridional entropy gradients in the convective layer
are nearly invariant in the vertical. Close to the equator,
large-scale vertical entropy fluxes lead to a weakly stable thermal
stratification above $\about 2$~bar, consistent with Jupiter's
stratification along the Galileo probe descent path
\citep{Magalhaes02}. The external gravity wave speed implied by the
thermal stratification\footnote{\label{fn:c}%
  We estimated the gravity wave speed as $c = \int_{p_t}^{p_s} N_p \,
  dp$, where $p_s$ is the pressure at the lower boundary,
  $p_t=0.15\,\mathrm{bar}$ is the upper boundary of the integration,
  $N_p^2 = -(\bar\rho \bar\theta)^{-1} \, \overline{\partial_p
    \theta}$ is a static stability measure, and $\rho$ is density
  \citep{Schneider06a}. Because the thermal stratification is (nearly)
  neutral in lower layers, the gravity wave speed depends only weakly
  on the lower boundary of the integration. However, it decreases as
  the upper boundary is lowered ($p_t$ is increased) within the stably
  stratified upper troposphere. For example, with an upper boundary of
  $p_t = 0.2\, \mathrm{bar}$, one obtains a gravity wave speed of
  $\about 350\,\mathrm{m\,s^{-1}}$ at the equator.} up to $0.15$~bar
is $\about 480\,\mathrm{m\,s^{-1}}$ at the equator, decreases to
$\about 400\,\mathrm{m\,s^{-1}}$ at $5^\circ$ latitude, and from there
decreases gradually but not monotonically to $\about
300\,\mathrm{m\,s^{-1}}$ at $80^\circ$ latitude---values roughly
consistent with the estimated gravity wave speed for Jupiter
\citep{Ingersoll95}.

Consistent with prograde jets being baroclinically more unstable than
retrograde jets and with baroclinic eddy generation preferentially in
prograde jets, eddies in the simulation transport angular momentum
meridionally from retrograde into prograde off-equatorial jets in
layers above $\about 1$~bar
(Fig.~\ref{f:vertical_structure}a). Poleward of $20^\circ$, the
divergence of meridional eddy angular momentum fluxes in retrograde
jets reaches up to $170\,\mathrm{m^2 \,s^{-2}}$ and the convergence in
prograde jets up to $470\,\mathrm{m^2 \,s^{-2}}$, with extremal values
attained between 0.2 and 0.6~bar. In the equatorial region, apparently
owing to meridional radiation of convectively generated Rossby waves,
there is strong convergence of meridional eddy angular momentum fluxes
(up to $890\,\mathrm{m^2 \,s^{-2}}$) between 0.6 and 1.4~bar---below
and near the transition from the weakly or neutrally stratified lower
layers to the more stably stratified upper layers. There is weaker
divergence (up to $660\,\mathrm{m^2 \,s^{-2}}$) at higher and lower
levels, as well as immediately off the equator between 0.6 and 2~bar.

\begin{figure}[!htb]
  \centering\includegraphics{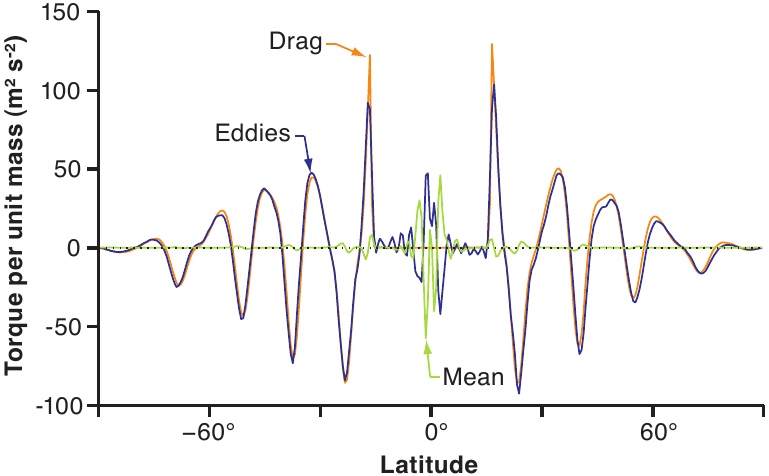}
  \caption{Zonally, temporally, and vertically averaged angular
    momentum balance in Jupiter simulation. Torque per unit mass owing
    to divergence of eddy angular momentum fluxes $\<\mathcal{S}_e\>$
    (blue), to divergence of mean angular momentum fluxes $\<
    \mathcal{S}_m\>$ (green), and to drag on the zonal flow $r_\perp
    \<\Bar{\mathcal{D}}\>$ (orange).}
  \label{f:am_balance}
\end{figure}
The angular momentum balance is closed in the manner discussed in
section~\ref{s:deep_flow}. Away from the equator, in the region with
Rayleigh drag, the Rossby number $\mathrm{Ro}=U/|fL|$ is small, and
mean meridional circulations link divergences and convergences of eddy
angular momentum fluxes in the upper troposphere to the drag at the
lower boundary. The mean meridional circulations satisfy the local
angular momentum balance \eqref{e:low_Rossby_am} in the upper
troposphere and extend downward along vertical lines (planetary
angular momentum contours in the thin-shell approximation); they form
thin-shell analogs of the circulations sketched in
Fig.~\ref{f:sketch}.  (In fact, we produced Fig.~\ref{f:sketch} by
projecting winds and mean meridional circulations from a thin-shell
GCM simulation into a thick spherical shell.)  Although the eddy
angular momentum fluxes are confined to a relatively thin layer, the
jets extend to the lower boundary, with the signs and zeros of the
zonal flow projecting downward approximately along vertical lines
(Fig.~\ref{f:vertical_structure}a). This allows the drag on the zonal
flow to balance any convergence or divergence of eddy angular momentum
fluxes in the vertical average, such that the vertically averaged
angular momentum balance in the limit of small Rossby number, $\<
\mathcal{S}_e \> \approx r_\perp \, \< \Bar{\mathcal{D}} \>$, is
approximately satisfied (Fig.~\ref{f:am_balance}).
In the region of vanishing Rayleigh drag, the vertically averaged
angular momentum balance $\<S_e + S_m\> \approx 0$ is satisfied by
partially compensating divergences and convergences of meridional eddy
angular momentum fluxes in different layers
(Fig.~\ref{f:vertical_structure}a), as well as vertical eddy fluxes
and mean meridional circulations redistributing angular momentum
between layers and between latitude bands within the equatorial region
(Fig.~\ref{f:am_balance}). The small residual in the zonally,
temporally, and vertically averaged angular momentum balance
(Fig.~\ref{f:am_balance}) is indicative of sampling variability and is
associated with low-frequency variability in the mean zonal flow.

The correlation between the convergence of eddy angular momentum
fluxes and the mean flow evident in Figs.~\ref{f:vertical_structure}
and \ref{f:am_balance} implies conversion of eddy kinetic energy to
mean flow kinetic energy. Above $1$~bar, the mean conversion rate from
eddy to mean flow kinetic energy per unit mass\footnote{We calculated
  energy conversion rates following \citet{Lorenz55}. The conversion
  rate from eddy to mean flow kinetic energy includes the (small)
  contributions owing to vertical eddy fluxes of angular momentum and
  other terms.} is $\about 2 \times 10^{-5}
\,\mathrm{W\,kg^{-1}}$. This conversion rate and the eddy angular
momentum fluxes themselves are of similar magnitude as those inferred
for Jupiter's upper troposphere by \citet{Ingersoll81} and
\citet{Salyk06}. However, unlike what was assumed in those studies,
the eddy angular momentum fluxes in our simulation have a baroclinic
structure and are confined to a relatively thin layer in the upper
troposphere. As a consequence, the total conversion rate from eddy to
mean flow kinetic energy in our simulation is only
$0.08\,\mathrm{W\,m^{-2}}$, that is, only 0.5\% of the energy uptake
by the atmosphere ($14.0\,\mathrm{W\,m^{-2}}$)---an order of magnitude
smaller than suggested by \citet{Ingersoll81} and \citet{Salyk06}. It
is also only a fraction (19\%) of the generation rate of eddy kinetic
energy by conversion from eddy potential energy
($0.41\,\mathrm{W\,m^{-2}}$). The remainder of the eddy kinetic energy
generation is balanced by dissipation (by the Rayleigh drag and by the
hyperdiffusion representing subgrid-scale processes), similarly as in
Earth's atmosphere \citep[cf.][chapter~14]{Peixoto92}.

\subsection{Eddy scales and turbulence characteristics}

Eddies in the simulation exhibit a broad range of length scales, with
the dominant scales depending on latitude and on whether zonal or
meridional velocity fluctuations are considered. 
\begin{figure}[!htb]
  \centering\includegraphics{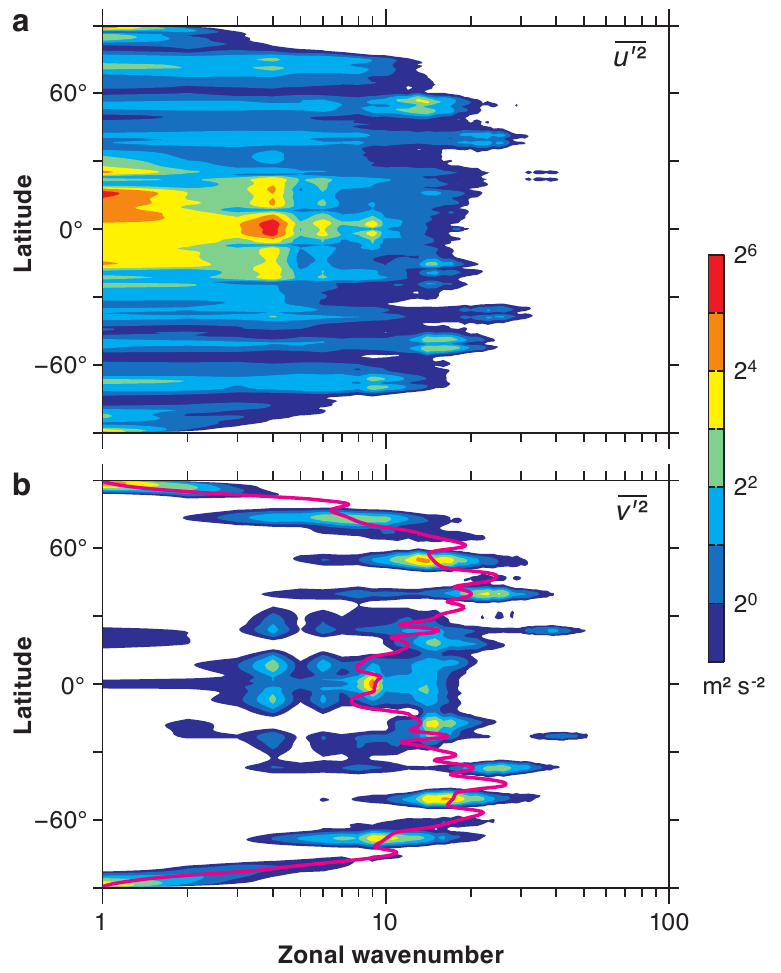}
  \caption{Mass-weighted vertical mean of zonal spectra of eddy
    velocity variances. (a) Zonal velocity variance
    $\overline{u'^2}$. (b) Meridional velocity variance
    $\overline{v'^2}$. The velocity variance contouring is
    logarithmic. The magenta line in (b) marks the energy-containing
    zonal wavenumber $m_e(\phi)$, defined as the first negative moment
    of the zonal spectrum of the meridional eddy velocity variance.}
  \label{f:zonspec}
\end{figure}
The vertically averaged zonal spectrum of the zonal eddy velocity
variance $\overline{u'^2}$ exhibits largest power over a broad range
of relatively small zonal wavenumbers $m \lesssim 10$
(Fig.~\ref{f:zonspec}a). The zonal eddy velocity variance is maximal
in the equatorial region, where it peaks at $m=4$ and retains
substantial power at yet smaller wavenumbers (see the equatorial waves
and wave packets in the instantaneous zonal velocity field in
Fig.~\ref{f:snapshots}d). In contrast, the analogous spectrum of the
meridional eddy velocity variance $\overline{v'^2}$ has relatively
well defined maxima at larger zonal wavenumbers $m \gtrsim 10$, except
in polar latitudes, where the maxima are at smaller wavenumbers
(Fig.~\ref{f:zonspec}b). At wavenumbers $m \gtrsim 10$ away from and
$m \gtrsim 5$ near the poles, zonal and meridional eddy velocity
variances are of similar magnitude, indicating approximate isotropy of
the eddies. At smaller wavenumbers, the zonal exceeds the meridional
eddy velocity variance, indicating that the zonal variance is
associated with anisotropic, predominantly zonal flow structures
\citep{Boer83b,Shepherd87a}. This large-scale zonal variance is likely
associated with variations in the zonal jets themselves, for example,
with low-frequency undulations. Therefore, the meridional eddy
velocity variance is more suitable for defining an energy-containing
turbulent eddy scale.

The energy-containing zonal wavenumber $m_e(\phi)$, defined as the first
negative moment of the vertically averaged zonal spectrum of the
meridional eddy velocity variance,\footnote{That is, given the
  vertically averaged meridional velocity variance spectrum $E_v(m,
  \phi)$, the energy-containing wavenumber is defined as the integral
  scale
  \[
  m_e(\phi) = \left( \frac{\sum_{m \ge 1} m^{-1} E_v(m, \phi)}{\sum_{m
        \ge 1} E_v(m, \phi)} \right)^{-1}.
  \]
  This integral scale more closely follows the maximum of the zonal
  spectrum than do other low-order moments of the spectrum.}  closely
follows the maximum of the zonal spectrum (Fig.~\ref{f:zonspec}b). It
increases from $\about 9$ near the equator to $\about 25$ near
$40^\circ$ latitude and then decreases toward the poles. The
associated zonal length scale $2\pi a\cos\phi/m_e$ decreases from
$\about 50000$~km near the equator to $\about 18000$~km near
$40^\circ$ latitude but then varies only weakly, between $\about
10000$~km and $\about 20000$~km, in higher latitudes. Poleward of
$\about 30^\circ$ latitude, this zonal length scale is similar to the
meridional separation scale between the off-equatorial jets
(Fig.~\ref{f:winds}a), which suggests that the length scale of
approximately isotropic eddies controls the jet spacing there.

The change in the energy-containing zonal wavenumber and length scale
between equatorial and off-equatorial regions marks a transition
between different dynamical regimes:
\begin{figure}[!htb]
  \centering\includegraphics{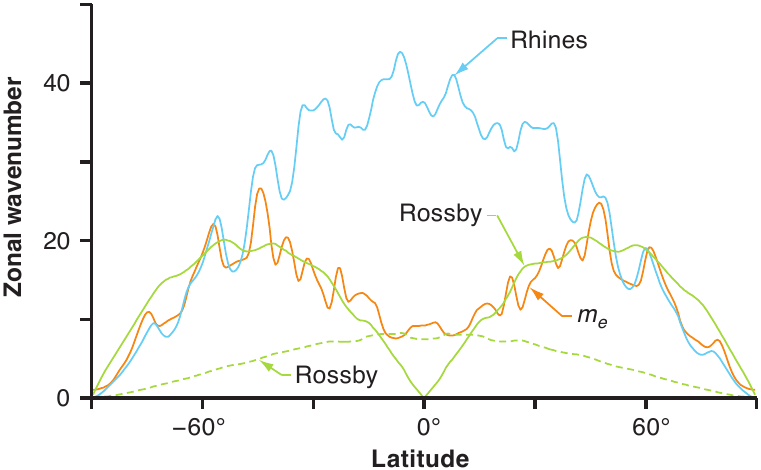}
  \caption{Zonal wavenumbers in Jupiter simulation. Orange:
    energy-containing wavenumber $m_e$. Light blue: Rhines wavenumber
    $a \cos\phi/L_\beta$ with $L_\beta = \gamma_\beta \,
    \mathrm{EKE_{bt}}^{1/4}/\beta^{1/2}$, barotropic eddy kinetic
    energy per unit mass $\mathrm{EKE_{bt}}$, and empirical constant
    $\gamma_\beta = 1.6$. Green dashed: equatorial Rossby wavenumber
    $a \cos\phi/L_e$ with $L_e = (c/\beta)^{1/2}$. Green solid:
    extratropical Rossby wavenumber $a \cos\phi/L_x$ with $L_x =
    \gamma_x c/|f|$ and empirical constant $\gamma_x = 1.8$. The
    gravity wave speed $c$ entering the Rossby wavenumbers is
    estimated from the thermal stratification between the lower
    boundary and $0.15$~bar (see footnote~\ref{fn:c}).}
  \label{f:eddy_scales}
\end{figure}
\begin{itemize} 
\item In the equatorial region, the energy-containing wavenumber
  ($\about 9$) is approximately equal to the wavenumber corresponding
  to the equatorial Rossby radius $L_e = (c/\beta)^{1/2}$
  (Fig.~\ref{f:eddy_scales}).\footnote{For an approximately isotropic
    equatorial Rossby wave with meridional decay scale $L_e$, we
    assume the zonal wavelength is $2\pi L_e$, and hence the zonal
    wavenumber is $2\pi a\cos\phi/(2\pi L_e) = a\cos\phi/L_e$
    \citep[cf.][]{Matsuno66}. Other zonal wavenumbers in what follows,
    in which $O(1)$ constants in the relevant length scales $L$ can be
    adjusted in any case, are defined analogously as $a\cos\phi/L$,
    that is, without factors of $2\pi$.}  Thus, it is approximately
  equal to the wavenumber of long equatorial Rossby waves.
  Consistently, the zonal phase velocity spectrum \citep{Randel91} of
  the meridional eddy velocity variance exhibits maximum power at
  retrograde phase velocities between $-50$ and
  $-100\,\mathrm{m\,s^{-1}}$, and the analogous spectrum of the
  meridional eddy flux of angular momentum peaks at similar phase
  velocities.\footnote{Long waves with retrograde phase velocities are
    also clearly evident in the zonal velocity animations available at
    www.gps.caltech.edu/$\sim$tapio/papers/.}  This is in agreement
  with theory: long equatorial Rossby waves are expected to have
  retrograde phase velocities of about $-c/3 \sim -150
  \,\mathrm{m\,s^{-1}}$ relative to the mean zonal flow
  \citep[chapter~11.8]{Matsuno66,Gill82}.
\item Away from the equator, the energy-containing wavenumber is
  approximately equal to the wavenumber corresponding to the
  extratropical Rossby radius $L_x = \gamma_x c/|f|$, where we fixed
  the empirical constant at $\gamma_x = 1.8$
  (Fig.~\ref{f:eddy_scales}). Thus, it scales approximately with the
  expected wavenumber of the baroclinically most unstable linear
  waves. In middle and high latitudes, the energy-containing
  wavenumber is also closely approximated by the wavenumber
  corresponding to the Rhines scale $L_\beta = \gamma_\beta
  \mathrm{EKE_{bt}}^{1/4}/\beta^{1/2}$, obtained by combining the
  barotropic eddy kinetic energy per unit mass $\mathrm{EKE_{bt}}$
  (the eddy kinetic energy of the vertically averaged flow) with
  $\beta$ \citep{Rhines75,Rhines94}; we fixed the empirical constant
  at $\gamma_\beta = 1.6$ as in \citet{OGorman08a}. The Rhines
  wavenumber even captures the variations in the energy-containing
  wavenumber between retrograde and prograde jets; however,
  equatorward of $\about 45^\circ$ latitude, the Rhines wavenumber
  exceeds the energy-containing wavenumber (Fig.~\ref{f:eddy_scales}).
\end{itemize}
These results are consistent with convectively generated equatorial
Rossby waves dominating the meridional eddy velocity variance in the
equatorial region, and baroclinically generated eddies dominating away
from the equator.

\begin{figure}[!htb]
  \centering\includegraphics{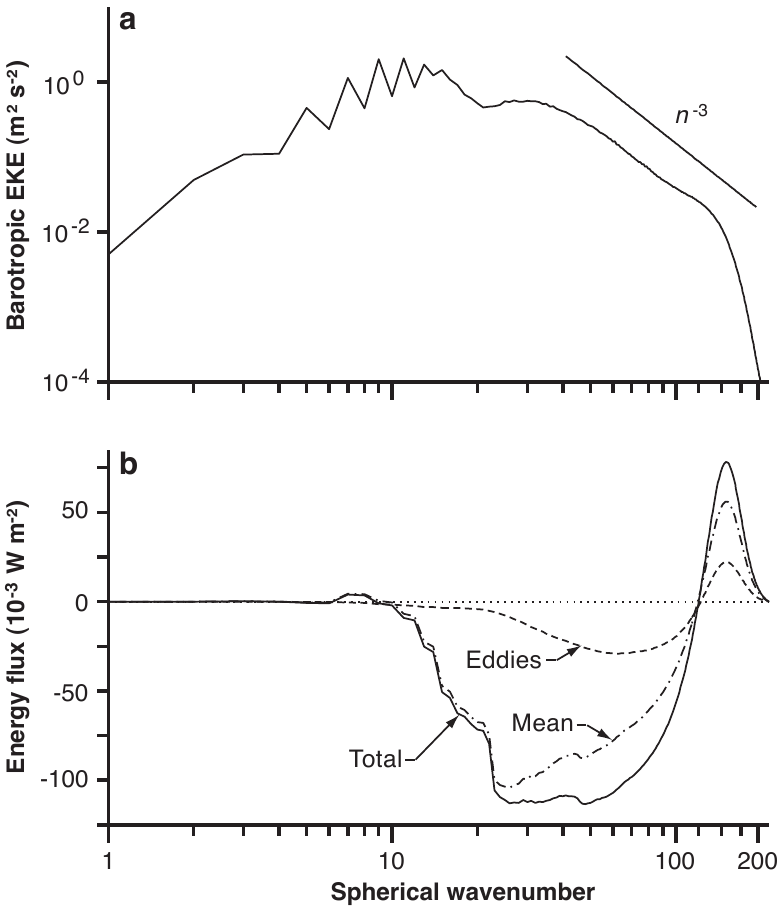}
  \caption{Eddy kinetic energy spectrum and spectral energy fluxes as
    a function of spherical wavenumber. (a) Spectrum of barotropic
    eddy kinetic energy (eddy kinetic energy of vertically averaged
    flow), with an $n^{-3}$ power law in spherical wavenumber $n$ for
    comparison. (b) Nonlinear spectral flux of total (globally
    integrated) kinetic energy (solid), decomposed into components
    involving interactions with the zonal mean (dash-dotted) and
    involving only eddy-eddy interactions (dashed). Positive fluxes
    indicate downscale and negative fluxes upscale energy
    transfer. The spectrum and spectral fluxes are calculated
    following \citet{Boer83a} and \citet{Shepherd87a}.}
  \label{f:sphere_spec}
\end{figure}
The coincidence of the energy-containing scale with the Rossby radius
and the jet separation scale away from the equator indicates that the
off-equatorial jets form without an inverse cascade of barotropic eddy
kinetic energy, or at least without an inverse energy cascade over an
extended inertial range. The fact that the energy-containing scale is
larger than the Rhines scale equatorward of $\about 45^\circ$ latitude
also points to the absence of an inverse energy cascade; the
coincidence of the energy-containing scale with the Rhines scale at
higher latitudes is no evidence to the contrary \citep{Schneider06a}.
Indeed, consistent with the absence of an inverse energy cascade, the
global barotropic eddy kinetic energy spectrum as a function of
spherical wavenumber $n$ does not show the flattening to an $n^{-5/3}$
power law at large scales that would be expected if an inverse energy
cascade were occurring (Fig.~\ref{f:sphere_spec}a). The nonlinear
spectral flux of total kinetic energy shows upscale transfer of
kinetic energy over a range of wavenumbers between $n\about 10$ and
$100$ (Fig.~\ref{f:sphere_spec}b), but it is dominated by interactions
that involve the zonal-mean flow (i.e., the $m=0$ component,
dash-dotted line in Fig.~\ref{f:sphere_spec}b).  Interactions that
involve only eddies (i.e., only zonal wavenumbers $m>0$, dashed line
in Fig.~\ref{f:sphere_spec}b) and that could give rise to an inverse
energy cascade are much weaker, and they do not exhibit an extended
wavenumber range in which their nonlinear spectral flux is constant,
as would be expected in an inertial range. It is evident that
eddy-mean flow interactions, in addition to or in place of nonlinear
eddy-eddy interactions, are fundamental for the formation of the
off-equatorial jets \citep{OGorman07,OGorman08a}. This situation is
similar to that in Earth's atmosphere, in which, likewise, eddy-mean
flow interactions are more important than nonlinear eddy-eddy
interactions for the upscale transfer of kinetic energy, and there is
no extended inertial range \citep{Shepherd87a,Shepherd87b}.

\section{Control simulations}\label{s:controls}

To demonstrate that, in our GCM, differential radiative heating is
indeed responsible for off-equatorial jets and intrinsic convective
heat fluxes for equatorial superrotation, we performed one simulation
without differential radiative heating and one without intrinsic heat
fluxes. In the simulation without differential radiative heating, an
intrinsic heat flux of $5.7\,\mathrm{W\,m^{-2}}$ is imposed at the
lower boundary, as in the Jupiter simulation, but insolation at the
top of the atmosphere is uniform and equal to the global-mean
insolation in the Jupiter simulation. In the simulation without
intrinsic heat fluxes, insolation at the top of the atmosphere varies
with the cosine of latitude, as in the Jupiter simulation, but no heat
flux is imposed at the lower boundary. Other parameters in these
control simulations are identical to those in the Jupiter simulation.

With the intrinsic heat flux but without differential radiative
heating, a prograde equatorial jet forms, flanked by two retrograde
jets; however, there are no off-equatorial jets
(Fig.~\ref{f:winds}b). With differential radiative heating but without
intrinsic heat fluxes, off-equatorial jets form with similar speeds
and widths as in the Jupiter simulation; however, the equatorial zonal
flow is retrograde (Fig.~\ref{f:winds}b).

\begin{figure*}[!tb]
  \centering\includegraphics{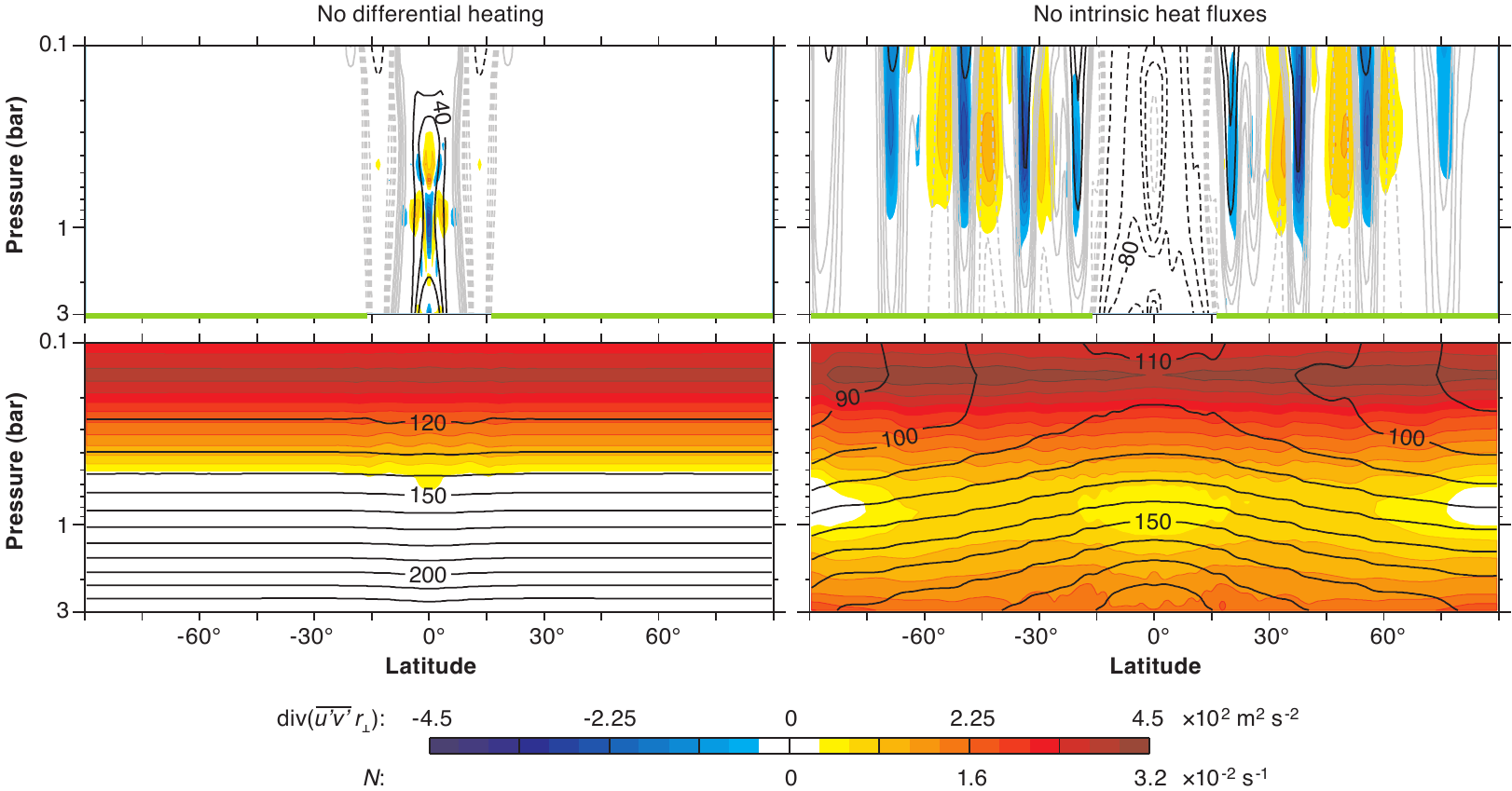}
  \caption{Flow fields in the latitude-pressure plane in control
    simulations. Top row: Zonal flow (contours) and divergence $\divy
    (\overline{u'v'} \, r_\perp)$ of meridional eddy angular momentum
    fluxes (colors). Bottom row: Temperature (contours) and
    Brunt-V{\"a}is{\"a}l{\"a} frequency $N$ (colors). Contour
    intervals and plotting conventions as in
    Fig.~\ref{f:vertical_structure}, except that the contour interval
    for the angular momentum flux divergence is halved. Left column:
    Simulation with intrinsic heat fluxes but with uniform insolation
    at the top of the atmosphere. Right column: Simulation with
    differential insolation but without intrinsic heat fluxes.  Shown
    are zonal and temporal means over the same 900 simulated days for
    which the zonal flows at 0.65~bar are shown in
    Fig.~\ref{f:winds}b.}
  \label{f:vertical_structure_ctrl}
\end{figure*}
The jets that do form in either case again extend to the lower
boundary, and the thermal structure is in thermal wind balance with
the zonal flow. In the simulation without differential radiative
heating (Fig.~\ref{f:vertical_structure_ctrl}, left column), as in the
Jupiter simulation, the thermal structure shows the signature of
convection penetrating into the upper troposphere, with stable
stratification above $\about 0.5\,\mathrm{bar}$ and nearly neutral
stratification beneath. The speed of the prograde equatorial jet again
increases with depth, albeit more weakly than in the Jupiter
simulation (cf.\ Fig.~\ref{f:vertical_structure}a). The speed of the
equatorial jet is smaller than in the Jupiter simulation, probably
because the control simulation lacks disturbances originating outside
the equatorial region, which may trigger convective heating
fluctuations and/or may interact with the equatorial mean zonal flow
in the Jupiter simulation. The atmosphere outside the equatorial
region is essentially in radiative-convective equilibrium. This
illustrates that convection does not necessarily generate kinetic
energy on large scales, although the convective heating in the
simulation fluctuates at all latitudes \citep[cf.][]{Emanuel94b}. In
the equatorial region, the weakness of horizontal temperature
gradients implies that convective heating fluctuations cannot generate
substantial temperature fluctuations but instead, if externally
forced, immediately generate large-scale horizontal divergence and
vorticity fluctuations and thus large-scale kinetic energy
\citep{Sobel01}. But outside the equatorial region, convective heating
fluctuations can generate local temperature fluctuations, which can
decay through radiative processes and dispersion by small-scale flows
that eventually dissipate, without generating substantial large-scale
kinetic energy. The assumption commonly made in shallow-water models
of Jupiter's upper atmosphere---that convective heating fluctuations
even away from the equator directly generate large-scale kinetic
energy---is in need of justification.

In the simulation without intrinsic heat fluxes
(Fig.~\ref{f:vertical_structure_ctrl}, right column), most of the
atmosphere is stably stratified, except for a nearly neutral layer
around $\about 0.8$~bar in high latitudes. The flow throughout the
equatorial region is retrograde. The speed of the off-equatorial zonal
flow decreases with depth in prograde jets and increases, with weaker
vertical shear, in retrograde jets, similarly as in the Jupiter
simulation (cf.\ Fig.~\ref{f:vertical_structure}a). The prograde
vertical shear of the zonal flow and, with it, equatorward temperature
gradients along isobars are generally larger in the simulation without
intrinsic heat fluxes, apparently because the baroclinicity of the
stably stratified atmosphere in the control simulation is smaller than
that of the atmosphere in the Jupiter simulation with a nearly
neutrally stratified layer; the smaller baroclinicity implies weaker
meridional eddy entropy fluxes and thus larger meridional temperature
gradients along isobars \citep[cf.][]{Schneider08b}. Consistent with
preferential eddy generation by baroclinic instability in prograde
jets, there is again meridional eddy angular momentum transport from
retrograde into prograde off-equatorial jets in layers above $\about
1$~bar.

The control simulations confirm that, in our GCM, both differential
radiative heating and intrinsic convective heat fluxes are necessary
to reproduce Jupiter's observed jets and thermal structure. The
imposed intrinsic heat flux needs to exceed a threshold value so that
convection penetrates into the upper troposphere. Calculations of
radiative equilibrium states show that, with the representation of
radiative processes in our GCM, the intrinsic heat flux needs to
exceed $\about 2\,\mathrm{W\,m^{-2}}$ for the troposphere to become
statically unstable. Indeed, the zonal flow in a simulation in which
an intrinsic heat flux of $2\,\mathrm{W\,m^{-2}}$ is imposed is
similar to that in the control simulation without intrinsic heat
fluxes.

\section{Conclusions and implications}

Based on the theory and simulations and consistent with the available
observational data, we propose that baroclinic eddies generated by
differential radiative heating are responsible for Jupiter's
off-equatorial jets, and that Rossby waves generated by intrinsic
convective heat fluxes are responsible for the equatorial
superrotation. Mean meridional circulations adjust entropy gradients
and the zonal flow in lower layers of Jupiter's atmosphere, such that
the zonal flow is in thermal wind balance with the entropy gradients,
and convergence or divergence of angular momentum fluxes in the upper
troposphere and the MHD drag on the zonal flow at depth balance upon
averaging over cylinders concentric with the planet's spin axis. As
demonstrated by the Jupiter simulation, the resulting view of how the
zonal flow and general circulation are generated and maintained is
consistent with observed large-scale features of Jupiter's jets and
thermal structure, such as the zonal flow and meridional temperature
variations in the upper troposphere and the thermal stratification of
the upper troposphere and layers beneath. It is also consistent with
the observed eddy angular momentum fluxes and with energetic
constraints indicating that these fluxes are confined to a relatively
thin atmospheric layer. As demonstrated by the control simulations,
differential radiative heating alone can account for the
off-equatorial jets, and intrinsic convective heat fluxes can account
for the prograde equatorial jet. However, intrinsic convective heat
fluxes alone do not necessarily lead to formation of off-equatorial
jets.

The theory and simulations predict aspects of the general circulation
that have not been observed but that are or soon will be
observable. For example, the transition in energy-containing eddy
scale between the equatorial region and regions away from the equator
(Fig.~\ref{f:eddy_scales}), pointing to different mechanisms of eddy
generation, should be observable by tracking cloud features. And we
predict that the measurements of NASA's upcoming Juno mission to
Jupiter will be consistent with zonal jets that extend deeply into the
atmosphere at all latitudes, away from the equator up to depths at
which the MHD drag acts. As already observed in the upper troposphere,
we expect that also at lower layers the speeds of the prograde
off-equatorial jets decrease with depth, and that there are associated
equatorward temperature gradients along isobars: qualitatively as in
Fig.~\ref{f:vertical_structure}a and b, but likely not quantitatively
so because the jets are expected to extend to much greater depth than
in our simulations and thus likely have weaker shear. Depending on the
strength of the MHD drag and on the depth at which it acts, the speeds
of the retrograde off-equatorial jets may decrease or increase with
depth, with weaker shear than the prograde jets, and with associated
poleward (reversed) or weaker equatorward temperature gradients. If
the shear of the zonal flow can be inferred from measurements, it can
be used together with observations of the eddy angular momentum
transport and with the implied transfer of kinetic energy from eddies
to the mean flow to constrain the strength and depth of the MHD drag
and thus to elucidate dynamics of the deep atmosphere that are not
amenable to direct measurement.

The proposed mechanisms are generic and likely act in the atmospheres
of all giant planets. They suggest, for example, that the reason that
Saturn's prograde equatorial jet is wider and stronger than Jupiter's
may be that Saturn's tropospheric gravity wave speed and equatorial
Rossby radius are greater. The greater depth at which MHD drag on
Saturn is estimated to act implies a wider region over which there is
effectively no drag on the zonal flow \citep{Liu08}, thus making a
wider equatorial jet possible. The proposed mechanisms also suggest
that the reason Uranus and Neptune do not exhibit equatorial
superrotation may be that their intrinsic heat fluxes are not
sufficiently strong to lead to convection penetrating into the upper
troposphere.

\paragraph{Acknowledgments}
This work was supported by a David and Lucile Packard Fellowship. The
GCM is based on the Flexible Modeling System of the Geophysical Fluid
Dynamics Laboratory; the simulations were performed on Caltech's
Division of Geological and Planetary Sciences Dell cluster. We thank
Isaac Held, Andrew Ingersoll, Yohai Kaspi, Paul O'Gorman, Adam Sobel,
David Stevenson, and Paul Wennberg for comments on drafts of this
paper and Paul O'Gorman for providing code for the calculation of
spectral energy fluxes.


\appendix{A}
\begin{center}
  \textbf{Average Angular Momentum Balance and Rossby Numbers}
\end{center}
\medskip\nopagebreak%

\nopagebreak
An average angular momentum balance follows from the balance equation
\begin{equation}\label{e:am_balance}
  \partial_t(\rho M) + \div (\rho \vec{u} M) = -\partial_\lambda p
  + r_\perp \rho \mathcal{D}
\end{equation}
for the angular momentum per unit mass about the planet's spin axis,
$M$, with longitude (azimuth) $\lambda$, three-dimensional velocity
vector $\vec{u}$, distance to the planet's spin axis $r_\perp$, and
drag force per unit mass on the zonal flow, $\mathcal{D}$. The angular
momentum per unit mass $M = M_{\Omega} + M_u$ consists of a planetary
component $M_\Omega = \Omega r_\perp^2$ owing to the planetary
rotation and a relative component $M_u = u r_\perp$ owing to the
relative zonal velocity $u$ of the atmosphere
\citep[e.g.,][chapter~11]{Peixoto92}. In this form, the angular
momentum balance \eqref{e:am_balance} is exact (up to neglected
viscous stresses) and holds irrespective of the constitutive laws of
the atmosphere. Averaging it temporally and zonally (over azimuth and
at constant $r_\perp$) gives, in a statistically steady state,
\begin{equation}\label{e:am_balance_loc_exact}
  \div (\overline{\rho \vec{u} M_u} + \overline{\rho \vec{u}}\,
  M_\Omega) = r_\perp \, \overline{\rho \mathcal{D}}, 
\end{equation}
where the overbar denotes a zonal and temporal mean. This is the
general, coordinate-independent form of the zonally and temporally
averaged angular momentum balance \eqref{e:am_balance_loc} in pressure
coordinates in section~\ref{s:superrotation}. If one additionally
averages \eqref{e:am_balance_loc_exact} along surfaces of constant
$M_\Omega$ (i.e., over lines of constant $r_\perp$ in the meridional
plane) and uses that, in a statistically steady state, there can be no
net mass flux across such fixed surfaces, it follows that only the
zonally and temporally averaged relative angular momentum fluxes
$\overline{\rho \vec{u} M_u}$ across an $M_\Omega$ surface contribute
to any net convergence or divergence of angular momentum fluxes and
any net acceleration or deceleration of the zonal flow averaged over
the $M_\Omega$ surface. This acceleration or deceleration must be
balanced by drag somewhere on the $M_\Omega$ surface; that is,
\begin{equation}
  \div \bigl\{\overline{\rho \vec{u} M_u}\bigr\} = r_\perp \,
  \bigl\{\overline{\rho \mathcal{D}}\bigr\}, 
\end{equation}
where $\{\cdot\}$ denotes the mean along lines of constant $r_\perp$
in the meridional plane. This is the general, coordinate-independent
form of the angular momentum balance \eqref{e:am_balance_vert} in
section~\ref{s:deep_flow}, where the relative angular momentum flux
$\overline{\rho \vec{u} M_u}$ is decomposed into mean and eddy
components and the zonal and temporal mean is taken along isobars. In
the thin-shell approximation, $r_\perp = a \cos\phi$ depends only on
latitude $\phi$; the average over $M_\Omega$ surfaces is the usual
zonal and vertical average \citep[e.g.,][]{Vallis06}.  In a deep
atmosphere, $r_\perp = r \cos\phi$ is the actual distance to the
planet's spin axis; the average over $M_\Omega$ surfaces is an average
along cylinders concentric with the spin axis.

The relative magnitude of relative and planetary angular momentum
advection by Eulerian mean flows is indicated by Rossby numbers, with
the relevant Rossby number depending on the direction of the flows
considered. For meridional flows (i.e., flows along meridians on
spheres concentric with the planet's center), the ratio of relative to
planetary angular momentum advection scales as $\partial_y
M_u/\partial_y M_\Omega \sim U/|f L|$ because $\partial_y M_\Omega = -
r_\perp f$, where $y$ is the meridional coordinate (arc length) and $L
\ll r_\perp$ is the meridional length scale of zonal-flow variations;
thus, this ratio scales with the usual Rossby number $\mathrm{Ro} =
U/|fL|$. For cylindrically-radial flows (i.e., flows perpendicular to
the planet's spin axis), the ratio of relative to planetary angular
momentum advection scales as $\partial_{r_\perp}
M_u/\partial_{r_\perp} M_\Omega \sim U/(2 \Omega L_\perp)$ because
$\partial_{r_\perp} M_\Omega = 2\Omega r_\perp$, where $L_\perp \ll
r_\perp$ is the cylindrically-radial length scale of zonal-flow
variations; thus, this ratio scales with the Rossby number
$\mathrm{Ro}_\perp = U/(2 \Omega L_\perp)$.

\appendix{B}
\centerline{\textbf{Jupiter GCM}}
\nopagebreak
\bigskip\noindent%

The GCM is based on the Flexible Modeling System developed at NOAA's
Geophysical Fluid Dynamics Laboratory. It uses standard Jupiter
parameters (Table~\ref{table_jupiter}), except where noted, and
integrates the hydrostatic primitive equations for a dry ideal-gas
atmosphere in a thin spherical shell, with a stress-free upper
boundary at zero pressure and a lower boundary with constant
geopotential and with a mean pressure of $3$~bar.

\begin{table*}[!htb]
  \caption{Parameters in Jupiter GCM}\label{table_jupiter}
  \centering
  \begin{tabular}{lll}
    \hline\hline\\[-2ex]
    \multicolumn{1}{c}{Parameter, symbol} & \multicolumn{1}{c}{Value}
    & \multicolumn{1}{c}{Reference}\\[.25ex]
    \hline\\[-2ex]
    Planetary radius (at 3~bar), $a$& 
    $69.86 \times 10^{6} \, \mathrm{m}$  & \cite{Guillot99} \\
    Planetary angular velocity, $\Omega$ &
    $1.7587 \times 10^{-4} \, \mathrm{s}$ & \cite{Donivan69}\\
    Gravitational acceleration, $g$ &
    $26.0\, \mathrm{m \, s^{-2}}$  & \cite{Lodders98}\\
    Specific gas constant, $R$ &
    $3605.38 \, \mathrm{J \, kg^{-1} \, K^{-1}}$  & \cite{Lodders98} \\
    Adiabatic exponent, $\kappa$ & $2/7$\\
    Specific heat capacity, $c_p=R/\kappa$ &
    $12619.0 \, \mathrm{J \, kg^{-1} \, K^{-1}} $ \\
    Solar constant, $F_0$ &
    $50.7 \, \mathrm{W\, m^{-2}}$ & \cite{Levine77} \\
    Intrinsic heat flux &
    $5.7 \, \mathrm{W\, m^{-2}}$ &  \cite{Gierasch00} \\
    Bond albedo, $r_\infty$ &
    $0.343$ & \cite{Hanel81} \\
    Single-scattering albedo, $\tilde{\omega}$ & $0.8$ & \cite{Sromovsky02}\\
    \hline
  \end{tabular}
\end{table*}

\subsection{Discretization and resolution}

The primitive equations in Bourke's \citeyearpar{Bourke74}
vorticity-divergence form are discretized with the spectral transform
method in the horizontal, finite differences in the vertical, and with
semi-implicit time-differencing
\citep[e.g.,][chapter~7.6]{Durran99}. The horizontal spectral
resolution of the GCM is T213 (triangular truncation of the spherical
harmonics expansion at wavenumber 213), with $640 \times 320$
(longitude$\, \times \,$latitude) points on the Gaussian transform
grid. The vertical coordinate is $\sigma = p/p_s$ (pressure $p$
normalized by pressure at lower boundary $p_s$) and is discretized
with 30 equally spaced levels. With this vertical discretization, the
density varies by two orders of magnitude from the top to the bottom
of the domain.

\subsection{Radiative transfer}  

Radiative transfer is represented as that in a homogeneous gray
atmosphere, using the two-stream approximation. The top-of-atmosphere
(TOA) insolation is imposed as a perpetual equinox with no diurnal
cycle,
\begin{equation}
  F_\mathrm{TOA} = \frac{F_0}{\pi} \cos\phi\, ,  
\end{equation}
where $F_0 = 50.7\,\mathrm{W\,m^{-2}}$ is the solar constant
\citep{Levine77}.

The solar optical depth $\tau_s$ is linear in pressure to represent
scattering and absorption by a well-mixed absorber,
\begin{equation}
\tau_s = \tau_{s0} \left( \frac{p}{p_0} \right)\, , 
\end{equation}
where $\tau_{s0}$ is the solar optical depth at pressure $p_0$.  We
assume diffuse incidence of solar radiation at TOA. The solar flux $F$
for a semi-infinite scattering and absorbing atmosphere then is
\citep{Petty06}
\begin{equation}
F = F_\mathrm{TOA} (1-r_\infty) \exp{\left(- \Gamma \tau_s\right)}\, , 
\end{equation}
where 
\begin{equation}
  \Gamma = 2 \sqrt{1 - \tilde{\omega}}\, 
  \sqrt{1 - \tilde{\omega}\gamma} \, ,
\end{equation}
and the Bond albedo $r_\infty$ can be represented as
\begin{equation}
  r_\infty = \frac{\sqrt{1 - \tilde{\omega} \gamma} - 
    \sqrt{ 1 - \tilde{\omega}}}
  { \sqrt{1 - \tilde{\omega} \gamma} + 
    \sqrt{ 1 - \tilde{\omega}}  }\, . 
\end{equation}  
Here, $\tilde{\omega}$ is the single-scattering albedo and $\gamma$
the asymmetry factor, which we chose to be $\tilde{\omega} = 0.8$ and
$\gamma = 0.204$ to give Jupiter's Bond albedo of $r_\infty = 0.343$
\citep{Sromovsky02,Hanel81}. For the solar optical depth, we chose
$\tau_{s0} = 3$ at $p_0 = 3 \, \mathrm{bar}$ to give a radiative flux
$F$ qualitatively consistent with measurements along the Galileo probe
descent path in Jupiter, except in the region of hazes in the upper
atmosphere \citep{Sromovsky98}. With these parameters, the solar
radiative flux at the lower boundary of the GCM is less than $9\%$ of
the incident flux at the top of the atmosphere.

The thermal optical depth $\tau_l$ is quadratic in pressure to
represent collision-induced absorption,
\begin{equation}
  \tau_l = \tau_{l0} \left( \frac{p}{p_0} \right)^2 \, , 
\end{equation}
where $\tau_{l0}$ is the thermal optical depth at pressure $p_0$. We
chose $\tau_{l0}=80$ at $p_0 = 3 \, \mathrm{bar}$, again to give
radiative fluxes qualitatively consistent with measurements along the
Galileo probe descent path and to give a thermal emission level (where
$\tau_l \sim 1$) in the vicinity of $\about 0.4$~bar
\citep{Ingersoll90,Sromovsky98}.

At the lower boundary, energy conservation is imposed: at each grid
point, the upward thermal radiative flux is set equal to the sum of
the downward solar and thermal radiative fluxes.

\subsection{Intrinsic heat flux}

A spatially uniform and temporally constant heat flux is deposited in
the lowest layer of the GCM to mimic Jupiter's intrinsic heat flux. We
use a heat flux of $5.7 \, \mathrm{W \ m^{-2}}$ \citep{Gierasch00},
except in the control simulation without intrinsic heat fluxes.
 
\subsection{Convection scheme}

A quasi-equilibrium convection scheme represents (dry) convection. It
relaxes temperature profiles toward a convective profile with
adiabatic lapse rate $g/c_p \approx 2.1 \, \mathrm{K \ km^{-1}}$
whenever an air parcel lifted adiabatically from the lowest model
level has positive convective available potential energy
\citep{Schneider06a}. The convective relaxation time is 6~h, chosen as
roughly the time it takes a gravity wave with speed $c\approx
450\,\mathrm{m\,s^{-1}}$ to traverse the equatorial Rossby radius
$(c/\beta)^{1/2} \approx 9500\,\mathrm{km}$. The convection scheme
conserves enthalpy integrated over atmospheric columns; it does not
transport momentum. It can be viewed as a dry limit of the
Betts-Miller convection scheme \citep{Betts86a,Betts86b}.

\subsection{Drag at lower boundary}

In Jupiter's atmosphere, the conductivity of hydrogen increases with
depth, making a continuous transition from very low values in the
outer atmosphere to a constant value reached at the depth at which
hydrogen becomes metallic, which occurs at $\about 1.4 \,
\mathrm{Mbar}$ or $\about 0.84$ Jupiter radii
\citep{Nellis96,Liu08}. Where the atmosphere is electrically
conducting, the interaction of the magnetic field with the flow leads
to Ohmic dissipation and retards the flow \citep{Liu08}.

\begin{figure}[!htb]
  \centerline{\includegraphics{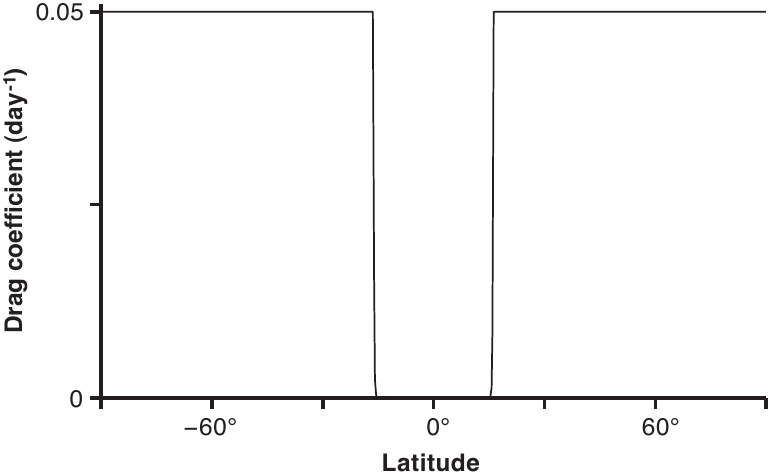}}
  \caption{Rayleigh drag coefficient $k_0(\phi)$ at the GCM's lower
    boundary.}
  \label{f:drag}
\end{figure}

As an idealized representation of effects of this MHD drag on the flow
in the outer atmosphere, we use Rayleigh drag near the GCM's lower
boundary in the horizontal momentum equations,
\begin{equation}
  \partial_t \vec{v} + \dots = -k(\phi,\sigma) \vec{v}.
\end{equation}
As in \citet{Held94}, the Rayleigh drag coefficient $k(\phi,\sigma)$
decreases linearly in $\sigma$ from its value $k_0(\phi)$ at the lower
boundary at $\sigma=1$ to zero at $\sigma_b = 0.8$,
\begin{equation}
  k(\phi,\sigma) = k_0(\phi) \max\left(0, \frac{\sigma - \sigma_b}{1-\sigma_b} \right).
\end{equation}
To represent in the thin-shell approximation the downward projection
along cylinders in a deep atmosphere, we use a drag coefficient
$k_0(\phi)$ that is constant ($k_0=0.05 \, \mathrm{day^{-1}}$)
poleward of $16.3^\circ$ latitude, corresponding to $0.96$ planetary
radii in a projection onto the equatorial plane, and that
exponentially decreases to zero at lower latitudes
(Fig.~\ref{f:drag}).

The kinetic energy dissipated by the Rayleigh drag is returned to the
flow as heat to conserve energy.

\subsection{Subgrid-scale dissipation}

Horizontal hyperdiffusion in the vorticity, divergence, and
temperature equations acts at all levels and is the only frictional
process above the layer with Rayleigh drag ($\sigma \le 0.8$). The
hyperdiffusion is represented by an exponential cutoff filter
\citep{Smith02b}, with a damping timescale of 2~h at the smallest
resolved scale and with no damping for spherical wavenumbers smaller
than 100.

\subsection{Simulations}

The simulations were spun-up from radiative-convective equilibrium
temperature profiles with no flow, with small random perturbations in
temperature and vorticity to break the axisymmetry of the initial
state. All simulations were first spun-up at T85 horizontal resolution
for 10000 simulated (Earth) days. The end states of the T85
simulations were used as initial states of T213 simulations, which
were spun up for at least 18000 additional days. Over the first
10000~days of the T213 simulations, we experimented with Rayleigh drag
parameters near the lower boundary to obtain a good fit to Jupiter's
observed zonal flow. The Rayleigh drag parameters were held fixed at
the values stated above for at least 8000~days of spin-up of the T213
simulations, until statistically steady states were reached.

In the statistically steady states, the global-mean outgoing thermal
radiative flux is within ${\lesssim 0.02\,\mathrm{W\,m^{-2}}}$ of the
sum of the global-mean absorbed solar radiative flux and the imposed
intrinsic heat flux. The vertically integrated Rayleigh drag on the
zonal flow approximately matches the vertically integrated total (mean
plus eddy) angular momentum flux convergence at each latitude
(Fig.~\ref{f:am_balance}).

The circulation statistics shown are computed from states sampled 4
times daily in the statistically steady states of the
simulations. Statistics for the Jupiter simulation are computed from
1500~simulated days; statistics for the control simulations are
computed from 900 simulated days. The statistics are first computed on
the GCM's $\sigma$ surfaces, with the appropriate surface
pressure-weighting of the averages \citep[e.g.,][]{Walker06}, and are
then interpolated to pressure surfaces for display purposes. The
divergence of eddy angular momentum fluxes is computed as the isobaric
divergence of the eddy fluxes interpolated to pressure surfaces; this
divergence does not differ significantly from the divergence of eddy
angular momentum fluxes on $\sigma$ surfaces.

\subsection{Sensitivity to Rayleigh drag coefficient}

If the Rayleigh drag coefficient $k_0(\phi)$ is taken to be constant
in latitude and equal to the value we use outside the equatorial
region ($0.05\, \mathrm{day}^{-1}$), much higher intrinsic heat fluxes
are needed to generate equatorial superrotation. In computationally
less demanding simulations of a planet with four times Earth's radius
but Jovian parameters otherwise (as in Table~\ref{table_jupiter}), we
found that the intrinsic heat fluxes had to be of order $100\,
\mathrm{W \, m^{-2}}$ to produce equatorial superrotation for a
constant drag coefficient of $0.05 \, \mathrm{day}^{-1}$.  If a
smaller constant drag coefficient is chosen ($\lesssim 0.001
\,\mathrm{day^{-1}}$), equatorial superrotation can be generated with
intrinsic heat fluxes comparable to $5.7 \, \mathrm{W\, m^{-2}}$, but
then the off-equatorial jets are much wider and stronger than
Jupiter's. The drag in the equatorial region has to be sufficiently
weak to produce equatorial superrotation with the observed intrinsic
heat flux, and the drag outside the equatorial region has to be
sufficiently strong to reproduce the speeds and widths of Jupiter's
off-equatorial jets.


\end{document}